# Single-sideband-interference twin-field quantum key distribution without global phase locking

Xingjian Li[1,6†], Bingkun Wang[1,6†], Jianyong Hu[1,2,6†*], Jianqiang Liu[3*], Shuxiao Wu[1,6], Guosheng Feng[4], Zhixing Qiao[4], Changgang Yang[1,6], Ruiyun Chen[1,6], Chengbing Qin[1,6], Guofeng Zhang[1,6], Liantuan Xiao[1,2,5,6*] and Suotang Jia[1,6]

[1]*State Key Laboratory of Quantum Optics Technologies and Devices, Institute of Laser Spectroscopy, Shanxi University, Taiyuan 030006, China*

[2]*Hefei National Laboratory, Hefei 230088, China*

[3]*College of Information Engineering, Shanxi Vocational University of Engineering Science and Technology, Jinzhong 030619, China*

[4]*College of Medical Imaging, Shanxi Medical University, Taiyuan, 030001, China*

[5]*College of Physics and Optoelectronics Engineering, Taiyuan University of Technology, Taiyuan 030024, China*

[6]*Collaborative Innovation Center of Extreme Optics, Shanxi University, Taiyuan 030006, China*

[†]*These authors contribute equally*

**Corresponding author E-mail address: jyhu@sxu.edu.cn; liujianqiang@sxgkd.edu.cn; xlt@sxu.edu.cn*

**Abstract:** Twin-field quantum key distribution (TF-QKD) can overcome the fundamental rate–loss limit of repeaterless quantum links, but its practical deployment has long been hindered by the requirement of global phase locking between two independent lasers. By revisiting the fundamental principles of optical interference, this work reveals that interference in TF-QKD inherently relies only on the instantaneous phase alignment of two independent optical pulses at the moment they temporally overlap, rather than on continuous global phase synchronization. Guided by this insight, we propose and demonstrate a single-sideband-interference TF-QKD protocol that eliminates global phase locking. Each user employs an I/Q modulator to generate a weak single sideband as the quantum signal, while the intrinsically phase-correlated optical carrier co-propagates as a real-time phase reference. Carrier interference at the receiver enables real-time phase extraction and feedback compensation for the sidebands. Unlike prior no-phase-locking approaches requiring second- or microsecond-level coherence, in principle, our scheme reduces this requirement to nanoseconds. We achieve 98% interference visibility over 100.8 km fibre and secure key rates surpassing the PLOB bound in the high-loss regime, providing a simpler route towards practical long-distance quantum communication networks.

## Introduction

Twin-field quantum key distribution (TF-QKD) has emerged as a landmark advancement in quantum communication [1]. By enabling the secret key rate to scale with the square root of channel transmittance, it offers a compelling solution to overcome the fundamental rate-loss bound that limits repeaterless quantum links [2, 3], thereby extending the secure communication distance well beyond the capabilities of conventional QKD schemes and holding particular promise for metropolitan and intercity quantum networks. Numerous experiments have demonstrated its feasibility over record distances [4-15]. Moreover, its measurement-device-independent architecture naturally supports star-topology networks, in which multiple users can securely establish keys via an untrusted central node, significantly reducing deployment complexity [16-18] and paving the way for deployed quantum networks [19-21].

Despite these advantages, the practical implementation of TF-QKD has long been obstructed by a stringent requirement for stable single-photon interference between two independent, remote laser. Prior efforts to ease this phase requirement have followed two distinct routes. The first, phase stabilization (global phase locking), relies on complex hardware such as optical phase-locked loops and ultrastable cavities to actively lock the global phase of two independent lasers, ensuring long-term coherence [10-15]. This approach is hardware-intensive and costly, and the time slots allocated for reference signals consume valuable quantum resources. The second, no-phase-locking operation (e.g., phase postselection or asynchronous pairing), relaxes the requirement by allowing later postselection of events where the relative phase happens to be stable [22-26]. However, these schemes still require global phase coherence on the second, and they either discard a large fraction of events or depend on ultrastable cavities to maintain long coherence.

In contrast, our work introduces a third paradigm: real-time phase extraction from a co-propagating carrier and feed-forward compensation. We revisit the fundamental principles of optical interference [27] and find that global phase locking is not necessary for interference between two independent optical pulses; rather, a well-defined relative phase within the temporal window of pulse overlap suffices. This insight shifts the paradigm from actively preventing phase drift to passively measuring it and compensating in real time, reducing the required phase coherence time from seconds to the nanosecond scale (the pulse overlap window) while eliminating the need for postselection. Guided by this insight, we propose and experimentally demonstrate a single-sideband-interference TF-QKD protocol that operates without global phase locking. At the transmitters, Alice and Bob each employ an I/Q modulator to generate a weak single sideband as the quantum signal, while the intrinsically phase-correlated optical carrier co-propagates as an instantaneous phase reference. The carrier and sideband experience the same channel perturbations and are separated at the receiver using optical filters. Interference between the carriers enables real-time extraction of the relative phase drift between the two users, which is then used for immediate feedback compensation of the sideband signals. The interference outcomes of the compensated sidebands are used for key generation. Our experimental results demonstrate that this scheme achieves a maximum single-sideband interference visibility of 98% over hundred-kilometer-scale fibre links and generates secure key rates that clearly surpass the repeaterless PLOB bound in the high-loss regime. By eliminating the need for complex active global phase-locking systems, this work provides a simpler, more robust, and scalable route towards practical long-distance quantum communication networks.

## SSB-TF-QKD Protocol

The core of the SSB-TF-QKD protocol lies in a shift of the phase-compensation paradigm from active prevention to real-time measurement and compensation. Rather than pursuing continuous global phase locking between independent lasers, the protocol directly compensates the instantaneous relative phase of the signal pulses. By exploiting the intrinsically strong phase correlation between the sideband and the optical carrier, a strong co-propagating carrier serves as an intrinsic phase reference, enabling real-time extraction and compensation of the relative phase drift between Alice and Bob, and thereby allowing stable long-distance interference between the sidebands.

**Sideband preparation and phase encoding.** The implementation of TF-QKD can be modelled as a tripartite quantum system consisting of two remote users, Alice and Bob, and a measurement node, Charlie. Alice and Bob each use an independent laser, and the corresponding optical fields can be written as follows:

$$E_A^{(0)}(t)=\varepsilon_0 e^{i[\omega_c t+\theta_A(t)]},\quad E_B^{(0)}(t)=\varepsilon_0 e^{i[\omega_c t+\theta_B(t)]}, \tag{1}$$

where $\omega_c$ is the angular frequency of the optical carrier, $\varepsilon_0$ is the field amplitude, and $\theta_{A/B}$(t) represents the instantaneous phase noise of the laser.

At the transmitter, Alice and Bob each use an I/Q modulator driven by a radio-frequency signal to generate a full-carrier single-sideband signal from a continuous-wave laser. In this scheme, the +1st-order sideband serves as the quantum signal for phase encoding of the random key bits (0 and $\pi$), while the co-propagating optical carrier is used as an intrinsic phase-reference pulse. Taking Alice as an example, let Ω denote the angular frequency of the RF drive signal and $\varphi_A$ its initial phase. Under proper bias conditions of the I/Q modulator, the output optical field can be approximated as the superposition of the carrier and a +1st-order sideband:

$$E_A^{SSB}(t)\approx\varepsilon_c e^{i[\omega_c t+\theta_A(t)]}+\varepsilon_s e^{i[(\omega_c+\Omega)t+\theta_A(t)+\varphi_A]}, \tag{2}$$

where $\varepsilon_c$ and $\varepsilon_s$ are the amplitudes of the carrier and the sideband, respectively. Equation (2) captures the core insight of our scheme. The phase of the sideband differs from that of the carrier by a fixed amount equal to the RF phase $\varphi_A$, i.e., $\varphi_{sideband}-\varphi_{carrier}=\varphi_A$. This relation is inherent to the modulation process and independent of the laser phase noise $\theta_A$(t), directly establishing the strong phase correlation between the carrier and the sideband. Bob's side follows the same principle, with the phase offset of his sideband relative to his carrier given by $\varphi_B$.

During the encoding stage, in each time slot, Alice and Bob independently select the encoding mode with probability $P_X$ or the decoy mode with probability (1 - $P_X$). In the encoding mode, the sender maps a random key bit $b \in \{0,1\}$ to an additional encoding phase $\varphi_b \in \{0, \pi\}$, making the sideband phase equal to $\varphi_{A/B}+\varphi_b$, and sets the mean photon number per pulse to $|\alpha|^2$. In the decoy mode, the sender randomly chooses an intensity from the set $\{\mu, \nu, \omega\}$ with $\omega \ll \nu < \mu$. To satisfy the security requirements of the decoy-state method, the sender also applies a uniformly random phase in $[0,2\pi)$ to the sideband of each pulse. The detailed derivation of the single-sideband modulation and the phase encoding principle is provided in Supplementary Note S1.

**Channel transmission and common-mode phase noise.** The continuous-wave light is then carved into optical pulses by an intensity modulator. After synchronization, the pulses are transmitted to Charlie through single-mode fibres of lengths $L_A$ and $L_B$, respectively. The fibre channels introduce phase drifts $\delta_A(t)$ and $\delta_B(t)$ due to environmental perturbations such as temperature fluctuations and acoustic vibrations. Because the frequency difference Ω between the carrier ($\omega_c$) and the sideband ($\omega_c$ + Ω) is much smaller than the optical frequency, the dispersion-induced delay difference between them is negligible, and the phase perturbations experienced by the two components during fibre propagation are highly common mode. Accordingly, the optical fields arriving at Charlie can be written as follows:

From Alice

$$E_A^C(t)\approx\begin{Bmatrix}\varepsilon_c e^{i[\omega_c t+\theta_A(t-\tau_A)+\delta_A(t)]}+\\ \varepsilon_s e^{i[(\omega_c+\Omega)t+\theta_A(t-\tau_A)+\delta_A(t)+\varphi_A+\varphi_b^A]}\end{Bmatrix}. \tag{3}$$

From Bob

$$E_B^C(t) \approx \begin{Bmatrix} \varepsilon_c e^{i\left[\omega_c t+\theta_B(t-\tau_B)+\delta_B(t)\right]} + \\ \varepsilon_s e^{i\left[(\omega_c+\Omega)t+\theta_B(t-\tau_B)+\delta_B(t)+\varphi_B+\varphi_b^B\right]} \end{Bmatrix}, \tag{4}$$

where $\tau_{A/B}=L_{A/B}n/c$ denotes the propagation delay, and $n$ is the refractive index of the fibre. The common-mode transmission characteristics are analyzed in Supplementary Note S2.

**Carrier phase extraction and real-time sideband phase compensation.** In the SSB-TF-QKD scheme, the purpose of phase compensation is to align the two remote sideband signals to a common phase reference frame for interference. This common reference frame is determined by the carrier phase of Alice (or, equivalently, by that of Bob). At Charlie's measurement node, the optical signals from the two users are interfered, after which the carrier and sideband are spatially separated using a narrowband optical filter, such as a fibre Bragg grating. The separated carriers are reflected, and their interference signals are measured by two single-photon detectors.

$$I_{carrier} \propto 1+V_C \cos\left(\Delta\Phi_{total}(t)\right), \tag{5}$$

where $V_C$ is the carrier interference visibility, and $\Delta\Phi_{total}(t)$ is the instantaneous relative phase difference between Alice and Bob, which includes both the intrinsic laser phase noise and the channel-induced phase perturbations.

$$\Delta\Phi_{total}(t)=\left[\theta_A(t-\tau_A)-\theta_B(t-\tau_B)\right]+\left[\delta_A(t)-\delta_B(t)\right], \tag{6}$$

here $\theta_A(t-\tau_A)$ - $\theta_B(t-\tau_B)$ denotes the phase difference arising from the free-running phase drift of the two independent lasers, while $\delta_A(t)$ - $\delta_B(t)$ represents the relative phase perturbation introduced by the fibre channels. By monitoring $I_{carrier}$, Charlie extracts the instantaneous relative phase difference $\Delta\Phi_{total}(t)$ and uses it to provide phase feedback for the sidebands, aiming to drive $\Delta\Phi_{total}(t)$ towards zero. A detailed characterization of the laser phase noise and the two-level feedback control is given in Supplementary Note S3. Because the carrier and the sideband experience nearly the same optical path before being separated, this feedback signal accurately represents the phase correction required for the sideband signals. It should be emphasized that this phase difference contains neither the encoding phase $\varphi_b$ nor the radio-frequency phase $\varphi_{A/B}$; therefore, the carrier phase does not leak any key information.

**Sideband interference and quantum measurement.** After phase compensation, the laser phase noise $\theta_{A/B}(t)$ and the channel-induced phase noise $\delta_{A/B}(t)$ are completely cancelled (Fig. 1c). Consequently, the relative phase between the two sideband signals used for interference becomes

$$\Delta\varphi_{sideband}=\left(\varphi_B-\varphi_A\right)+\left(\varphi_b^B-\varphi_b^A\right), \tag{7}$$

which contains only two fixed contributions, one is a constant phase offset $\Delta\varphi_{RF}=\varphi_B$ - $\varphi_A$ determined by the RF drive signals (set to zero in this work after system calibration), and the other is the encoding phase difference $\Delta\varphi_b=\varphi_b^B-\varphi_b^A\in\{0,\pi\}$ that carries the key information. Therefore, stable photon interference can in principle be achieved without requiring frequency or phase locking between the two independent lasers.

The photon-detection probabilities at the two output ports of the sideband interference are given by:

$$P(D_1)\propto 1+V_S\cos\left(\Delta\varphi_{sideband}\right), \tag{8}$$

$$P(D_2)\propto 1-V_S\cos\left(\Delta\varphi_{sideband}\right), \tag{9}$$

where $V_S$ is the sideband interference visibility. When $\Delta\varphi_b = 0$, the photon is more likely to be detected by $D_1$; when $\Delta\varphi_b = \pi$, the photon is more likely to be detected by $D_2$. After the measurement, Charlie announces a click event at either $D_1$ or $D_2$ through the public channel, thereby completing one measurement round.

**Security analysis and key rate.** The protocol operates within a measurement-device-independent security framework [16]. As an untrusted third party, Charlie publicly announces all measurement outcomes, including the carrier-phase information. Although the strong optical carrier is measured as a classical signal, it carries no key-bit information; therefore, publicly disclosing the extracted phase difference $\Delta\Phi_{total}(t)$ does not compromise key security.

The protocol employs the standard decoy-state method to counter photon-number-splitting attacks [28, 29]. Alice and Bob randomly transmit phase-randomized coherent states with mean photon numbers $\mu$, $\nu$ and $\omega$ (decoy states, with $\omega \ll \nu < \mu$). By publicly comparing the decoy-state data, one can estimate the gains $Q_{kl}$ and quantum bit error rates $E_{kl}$ for different intensity combinations in the channel. Finally, in the asymptotic limit, the secure key rate is given by:

$$R = q \cdot Q_{\alpha\alpha}^{X}\left[1 - H_2\left(e_{bit}^{X}\right) - H_2\left(e_{ph}^{U}\right)\right], \tag{10}$$

where $q$ is the sifting factor (typically 1/2 in this protocol); $Q_{\alpha\alpha}^{X}$ is the total gain in the encoding mode (with an average number of photons per pulse $|\alpha|^2$); $e_{bit}^{X}$ is the bit error rate; and $H_2(x) = -x\log_2(x) - (1-x)\log_2(1-x)$ denotes the binary Shannon entropy.

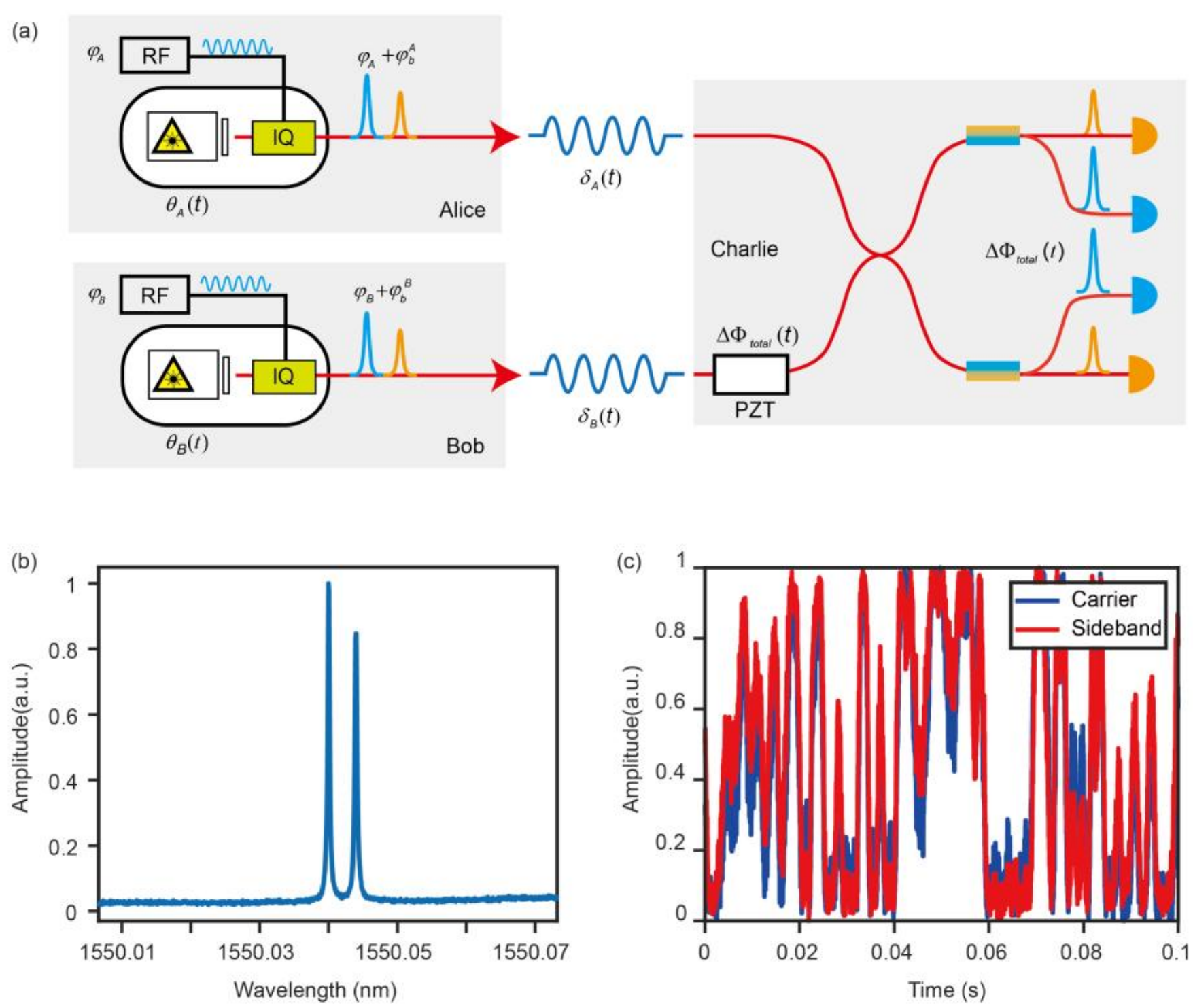

**Figure 1 | SSB-TF-QKD protocol. (a)** Schematic of the overall protocol architecture. Alice and Bob each use an I/Q modulator to generate a carrier and a single sideband. The sideband carries the encoding phase $\varphi_{A/B}+\varphi_b$, and the carrier serves as an intrinsic phase reference. After fibre transmission, the carrier and sideband are separated at Charlie by an optical filter. Interference of the carriers extracts the total phase difference $\Delta\Phi_{total}(t)$, which is fed back to compensate the sideband phase. Interference of the compensated sidebands then produces the key bits. **(b)** Typical optical spectrum of the single-sideband signal. **(c)** Measured interference patterns of the carrier and the sideband from two independent lasers.

The two traces exhibit highly consistent temporal variations, demonstrating that the carrier and sideband undergo the same random phase drifts and confirming their strong phase correlation, as well as the common-mode transmission property that enables the carrier to serve as an effective phase reference.

## Experimental Results and Analysis

We systematically investigated the feasibility and performance of the proposed SSB-TF-QKD protocol through a series of experiments (see Methods for detailed setup and procedures). The results provide a step-by-step validation of the scheme at four levels: single-sideband generation, phase correlation between the carrier and the sideband, real-time phase-compensation performance, and secure key generation. Collectively, these results demonstrate that the proposed protocol enables TF-QKD beyond the repeaterless quantum communication limit without global phase locking, offering a more practical route for long-distance quantum communication.

**Single-sideband generation.** High-quality single-sideband signals are essential for the protocol. By precisely controlling the DC bias of the I/Q modulator and the relative phase between the two RF drive signals, we achieved single-sideband modulation. As shown in the measured optical spectrum in Fig. 1b, the suppression ratio of the unwanted sideband (-1st order) is 30.34 dB, and that of the higher-order sidebands is 31.54 dB. These results confirm the clean generation of the target sideband (+1st order), establishing a solid basis for high-visibility interference.

**Verification of the phase correlation between the carrier and the sideband.** To validate the key physical insight of the proposed protocol, namely that the carrier and sideband undergo common-mode phase perturbations and can thus serve as intrinsic phase references, we synchronously measured the temporal evolution of the carrier and sideband interference signals after 100.8 km fibre transmission. In the experiment, the encoding phases of both Alice and Bob were set to zero, so that the observed phase variations arose solely from the intrinsic phase noise of the two independent lasers and the fibre-induced drift. As shown in Fig. 1c, the interference fringes of the carrier and sideband exhibit highly consistent temporal dynamics, with only a fixed offset between them. This result directly confirms the strong phase correlation predicted by the theoretical model and provides a solid experimental foundation for using carrier interference to extract and compensate channel phase noise in real time.

**Real-time phase compensation.** Using the phase reference extracted from carrier interference, we implemented a real-time phase-feedback system. At Charlie's receiver, the interfered optical signals were first spectrally separated by a fibre Bragg grating (FBG), which reflected the carrier while transmitting the sideband. The FBG provided an isolation of 32 dB, effectively suppressing carrier-photon-induced dark-count-like noise in sideband detection. Because fibre dispersion introduces a fixed relative delay between the carrier and the sideband, this constant phase offset was actively compensated by adjusting the RF phase shifters at the transmitters.

The carrier interference intensity was then monitored using two gated single-photon detectors. Photon counts were accumulated by an FPGA board to calculate the interference visibility, and a PID controller provided real-time feedback to a piezoelectric transducer-based fibre stretcher, thereby dynamically locking the system to the interference maximum, that is, maintaining $\Delta\Phi_{total}(t) \approx 0$. The performance of the phase-compensation system is directly reflected in the stability of the single-photon sideband interference. Figures 2a and 2b show the evolution of the photon count rate and the corresponding sideband interference visibility before and after feedback lock activation. Before locking, the interference signal fluctuated strongly and the visibility was undefined. After lock engagement, the photon counts rapidly stabilized, and the instantaneous visibility reached 98.0%. The long-term average visibility remained above 96.9% for over one hour. These results demonstrate that our phase-compensation system effectively converts the random phase noise from the lasers and fibre channel into a common-mode perturbation that is corrected in real time.

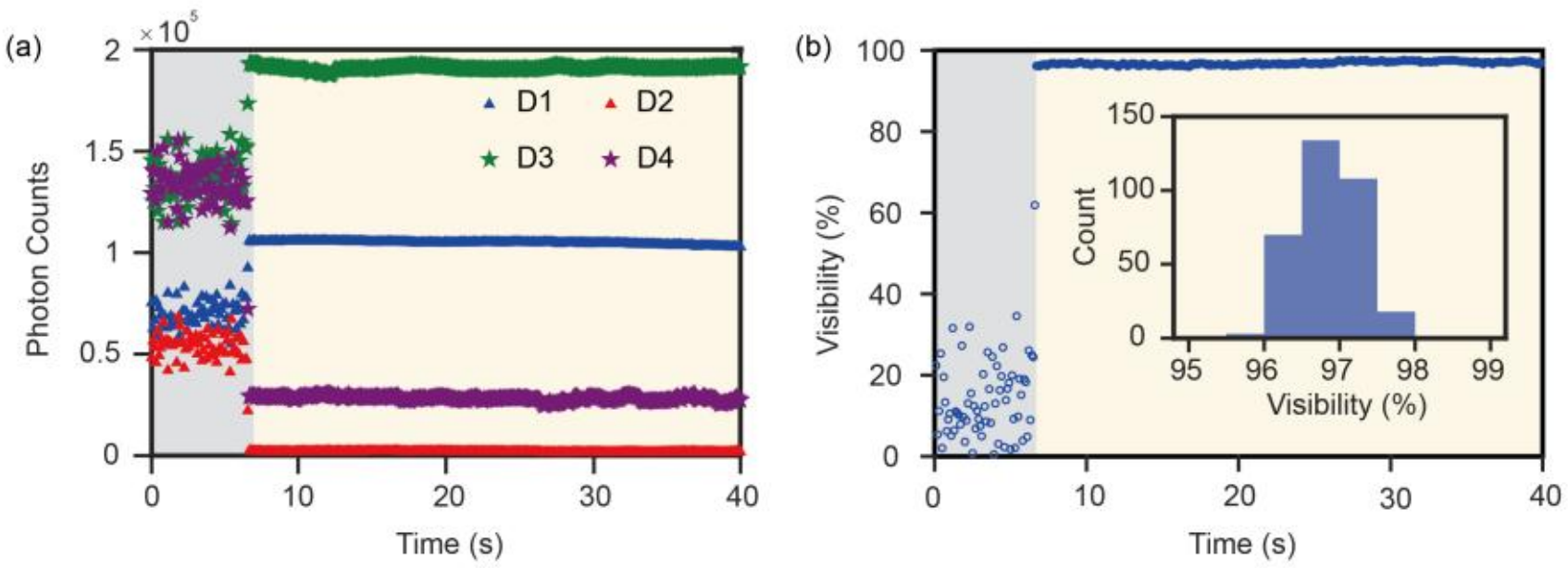

**Figure 2 | Performance of real-time phase compensation. (a)** Photon count rates for sideband interference (detectors D1 and D2) and carrier interference (D3 and D4) before and after feedback lock activation. The grey region indicates the unlocked state, where the interference signals fluctuate strongly and no stable fringes are established. The yellow region indicates the locked state, where the count rates at all four detectors rapidly stabilize, showing that the phase-feedback system suppresses dynamic phase noise to a negligible level. **(b)** Evolution of sideband interference visibility before and after lock activation. Before locking, the visibility is undefined. After lock engagement, the instantaneous visibility reaches 98.0%.

**Secure key generation.** Complete protocol cycles including state preparation, transmission, interference measurement, and post-processing were performed under three different total channel loss conditions of 33.7 dB, 42.9 dB, and 52.9 dB. In the experimental setup, the servo fibre link for coarse laser frequency alignment was fixed at 100.8 km to maintain baseline frequency stability. Different total losses were introduced by inserting a variable optical attenuator into the quantum channel. For the experiment with a total loss of 33.7 dB, the fibre link length was 0 km (0 dB), an additional variable attenuation of 10.1 dB was introduced, and the fixed insertion loss of the system, including devices and connections, was 23.6 dB. For the 42.9 dB total loss, the fibre link length was 100.8 km (19.3 dB), no additional attenuation was used, and the fixed insertion loss remained 23.6 dB. For the 52.9 dB total loss, the fibre link length was 100.8 km (19.3 dB), an additional variable attenuation of 10.0 dB was applied, and the fixed insertion loss was again 23.6 dB.

Table 1 | Optimized intensity settings, measured QBER, and secure key rates for various total channel losses.

| Loss (dB) | Fiber length (km) | | Intensities (counts/per pulse) | | | | QBER | | $R_{exp}$ | PLOB |
|---|---|---|---|---|---|---|---|---|---|---|
| Total | $L_S$ | $L_Q$ | $\|\alpha\|^2$ | $\mu$ | $\nu$ | $\omega$ | $D_0$ | $D_1$ | | |
| 33.7 | 100.8 | 0.0 | 0.0511 | 0.0858 | 0.0108 | 4.2064e-4 | 1.8518e-2 | 2.0408e-2 | 1.4307e-4 | 6.2989e-4 |
| 42.9 | 100.8 | 100.8 | 0.1189 | 0.1885 | 0.0466 | 1.3828e-3 | 2.4193e-2 | 2.6956e-2 | 8.9280e-5 | 7.3992e-5 |
| 52.9 | 100.8 | 100.8 | 0.0731 | 0.1107 | 0.0103 | 3.3071e-3 | 2.7522e-2 | 3.2258e-2 | 1.9754e-5 | 7.3990e-6 |

Note: Total denotes the total loss. $L_S$ the service-channel length, and $L_Q$ the quantum-signal link length. $|\alpha|^2$ is the mean photon number per pulse in the encoding mode, and $\{\mu, \nu, \omega\}$ are the mean photon numbers per pulse in the decoy mode. QBER denotes the quantum bit error rate, and $D_0$ and $D_1$ represent the error rates measured at the two detectors. The experimental key rate, $R_{exp}$, is the average over three independent experimental runs. PLOB denotes the repeaterless secret-key rate bound.

For each channel loss, we optimized the intensity set $\{|\alpha|^2, \mu, \nu, \omega\}$ for the signal and decoy states to maximize the secure key rate. The optimized values and the corresponding measured quantum bit error rates (QBERs) are listed in Table 1. All post-processing was performed within a rigorous measurement-device-independent security framework combined with decoy-state analysis (see Supplementary Note S5 for details). Specifically, the experimentally observed gains for different intensity combinations were used to estimate the single-photon component gain and an upper bound on the phase-error rate, from which the final secure key rate was derived.

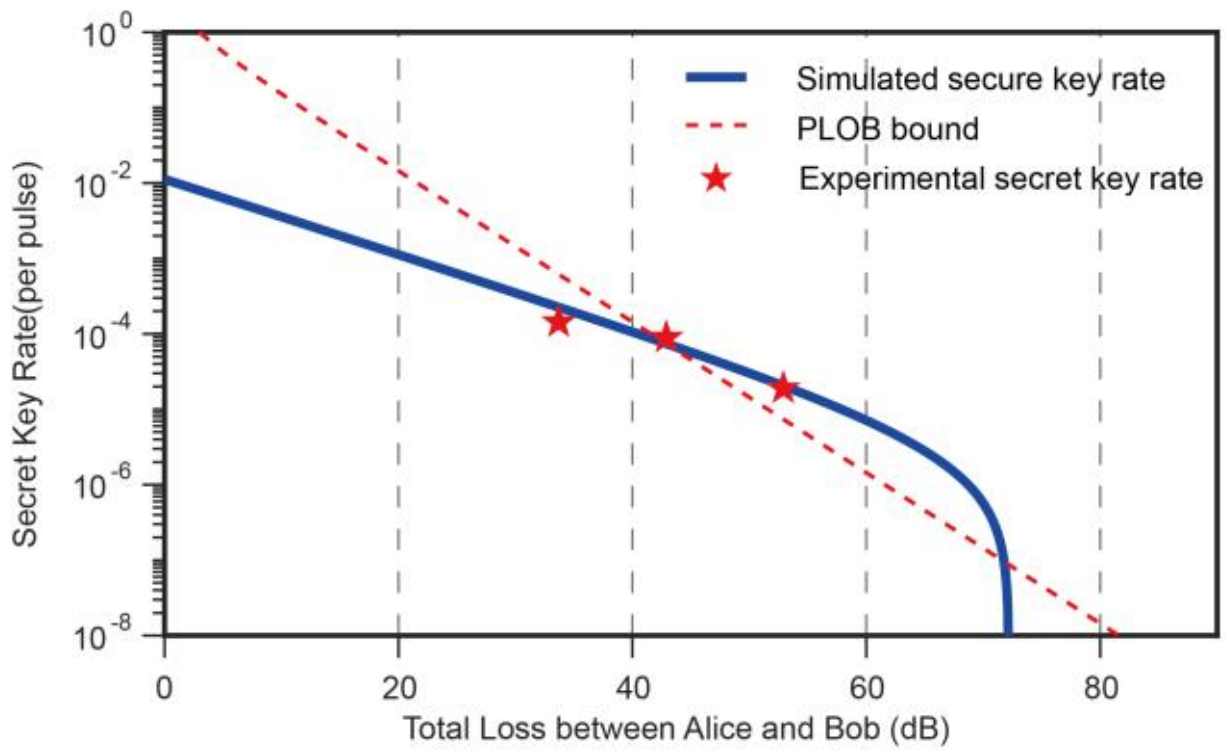

**Figure 3 | Secure key rate versus total channel loss.** The red dashed line represents the PLOB bound for repeaterless quantum communication, and the blue solid line is the theoretical simulation based on experimental parameters. The star symbols indicate the secure key rates obtained from three independent experiments. When the total channel loss exceeds 42.9 dB, the experimental secure key rate clearly surpasses the PLOB bound.

Figure 3 shows the experimental secure key rates as a function of total channel loss, together with the PLOB bound and the theoretical simulation. The experimental results show that, at a total channel loss of 42.9 dB, including a 100.8 km fibre link, the secure key rate reaches $8.93\times10^{-5}$ bit per pulse, which is approximately 1.2 times the PLOB bound. When the total channel loss increases to 52.9 dB, the secure key rate is $1.98\times10^{-5}$ bit per pulse, corresponding to approximately 2.7 times the PLOB bound, with the advantage becoming even more pronounced. The experimental gains and the estimated yields for different intensity combinations are listed in Supplementary Note S6. These results provide strong evidence that the proposed SSB-TF-QKD scheme can not only operate stably without global phase locking, but also surpass the repeaterless rate-loss limit. In particular, in the high-loss regime relevant to intercity communication (>40 dB), the experimental key rates clearly exceed the PLOB bound, establishing a solid experimental foundation for practical long-distance quantum-secure communication.

It is worth noting that the single-sideband modulation architecture adopted in this scheme offers good scalability. An I/Q modulator can simultaneously generate the positive and negative first-order sidebands, and their phases can be independently controlled through the phase difference of the RF drive signals and the modulator bias (see Supplementary Note S7 for details). Therefore, the present scheme can be naturally extended to a multi-sideband interference architecture. By encoding two orthogonal sidebands within a single optical pulse, the encoding efficiency can be doubled, thereby increasing the secure key rate under the same system resources and channel conditions. This extension preserves the core advantage of the scheme, namely the absence of any requirement for global phase locking, while providing a simple and effective route to improve the key generation rate.

## Conclusion

By re-examining the fundamental principles of optical interference, this work proposes a twin-field quantum key distribution protocol based on single-sideband interference that operates without global phase locking. In this protocol, an I/Q modulator generates a weak single sideband as the quantum signal while the intrinsically phase-correlated carrier co-propagates as a real-time phase reference. The carrier and sideband experience the same channel perturbations, enabling the carrier interference at the receiver to extract the instantaneous phase drift for immediate feedback compensation of the sideband signals. This design intrinsically immunizes the system against fibre-induced phase noise. Experimentally, the scheme achieves a maximum single-sideband interference visibility of 98% over a 100.8 km fibre link, with a long-term average visibility exceeding 96.9%, and generates secure key rates that surpass the repeaterless PLOB bound in the high-loss regime. Crucially, our scheme reduces the required phase coherence time from seconds to nanoseconds by exploiting the co-propagating carrier as a real-time phase reference. This shifts the paradigm from actively locking the global phase of two independent lasers to passively measuring and compensating the instantaneous phase drift in real time. Consequently, the need for ultrastable cavities, complex phase-locked loops, and dedicated reference frames is eliminated, dramatically lowering system complexity, cost, and environmental sensitivity. This advance makes TF-QKD a practical and robust candidate for deployment in standard telecom infrastructure.

## Methods

The experimental set-up of the SSB-TF-QKD system is shown in Fig. 4. It consists of two transmitter nodes, Alice and Bob, and a receiver node, Charlie. The transmitter nodes prepare the carrier-single-sideband optical pulses and perform phase encoding, while the receiver node carries out interference measurement, phase compensation.

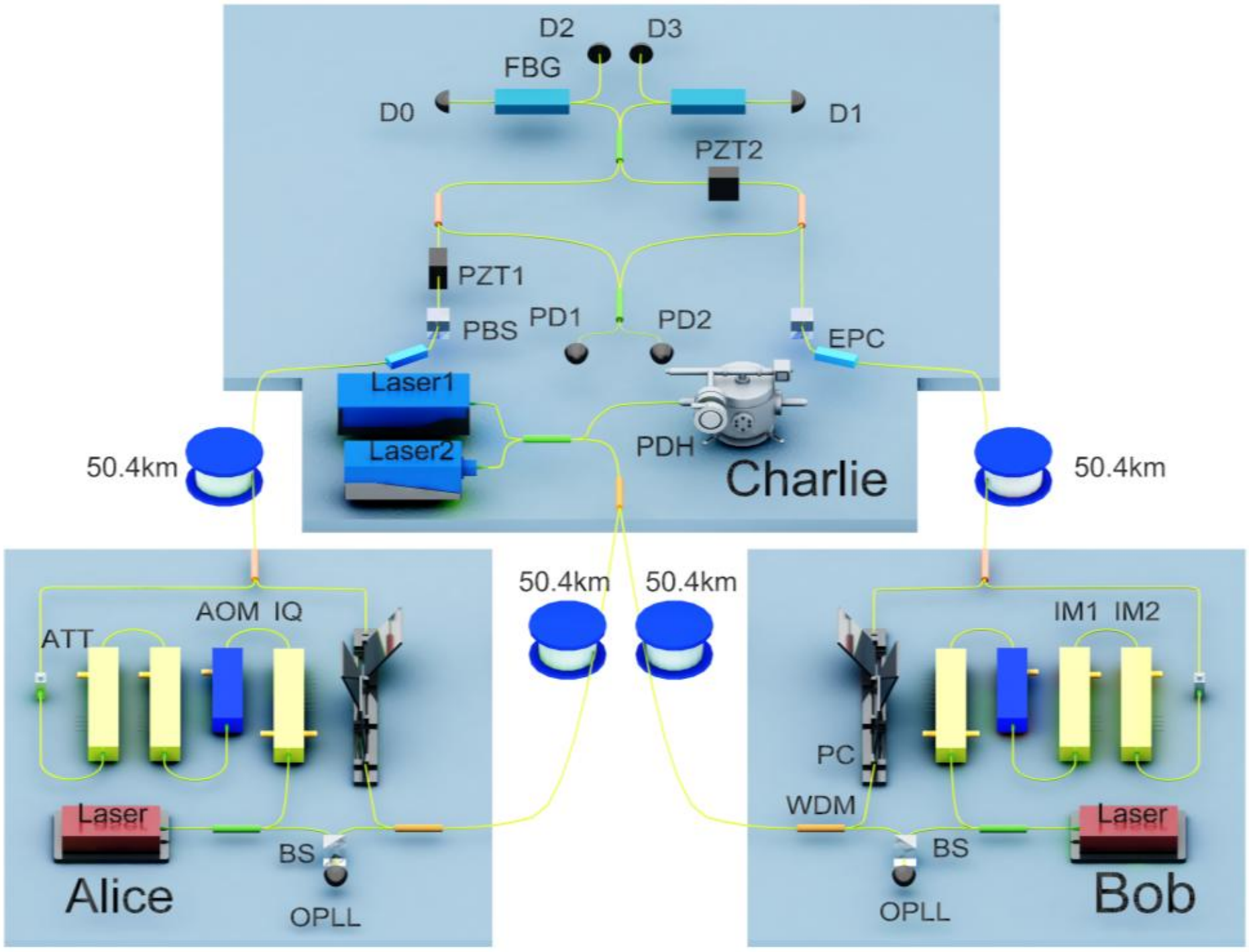

**Figure 4 | Experimental set-up of the SSB-TF-QKD system.** Lasers 1 and 2 at Charlie have wavelengths of 1550.04 nm and 1538.48 nm, respectively. Their outputs are wavelength-division multiplexed and transmitted to Alice and Bob through 50.4 km of single-mode fibre. At each transmitter, a WDM demultiplexes the two wavelengths. The light at 1550.04 nm is used to lock the user lasers. An acousto-optic modulator (AOM) is employed to align the frequencies of the two user lasers, and an intensity modulator (IM) is used for pulse carving of the signal light and decoy-state modulation. The signals from Alice and Bob are then sent to Charlie, where polarization fluctuations are compensated by

electrically controlled polarization controllers (EPCs) and polarization beam splitters (PBSs). After wavelength separation, the two optical channels are detected separately. Laser, laser source; OPLL, optical phase-locked loop; BS, beam splitter; PC, polarization controller; ATT, variable optical attenuator; PZT, piezoelectric transducer-based fibre stretcher; FBG, fibre Bragg grating.

**Optical field preparation and phase encoding.** At the transmitter, Alice and Bob use two independent distributed-feedback lasers as optical sources, with a central wavelength of 1550.04 nm and linewidths of 160 Hz and 224 Hz, respectively. To satisfy the coherence requirement that the coherence time be much longer than the pulse width, we implement an optical phase-locked loop. A reference laser located at Charlie is delivered to each transmitter through independent fibre links and interferes with the local laser, forming a phase-locked loop that suppresses laser frequency drift to the order of kHz per second. We stress that this frequency stabilization serves only to keep the frequency difference between the two independent lasers within a range where the coherence time exceeds the pulse width (1 ns). It does not lock their global phases. In a proof-of-principle experiment, this simplification allows us to focus on validating the core idea of eliminating global phase locking. For practical deployment, one could use lasers with intrinsic frequency stability sufficient for the required coherence time, e.g., those locked to atomic transitions, thereby removing even the need for active frequency stabilization.

The laser is then fed into an I/Q modulator for single-sideband modulation. The modulator is driven by a 5.5 GHz radio-frequency signal. After passing through a power splitter, digitally controlled phase shifters, and amplifiers, two drive signals with equal amplitude and a phase difference of $\pi/2$ are applied to the I and Q arms, respectively. By carefully setting the DC bias, we suppress the lower (-1st order) sideband and generate a full-carrier single-sideband signal consisting of the carrier and the upper sideband. The suppression ratio of the unwanted sideband (-1st order) is 30.34 dB, and that of the higher-order sidebands is 31.54 dB. The resulting continuous optical field then passes sequentially through two intensity modulators. The first intensity modulator carves the light into optical pulses at a repetition rate of 10 MHz with a pulse width of 1 ns. The second intensity modulator is used for decoy-state modulation. According to instructions from a random-number generator, the pulse intensity is randomly set to $\mu_i \in \{\mu, \nu, \omega\}$, so as to satisfy the requirements of the decoy-state method used in the security proof.

Phase encoding is implemented by controlling the phase of the RF signal. As shown in Fig. 4, the FPGA outputs a pseudorandom sequence at 10 MHz to control an RF switch, which randomly selects between two digitally controlled phase shifters. The phase shifters are preset to the encoding phases 0 and $\pi$, respectively, and the state of the RF switch determines the final phase $\varphi_{A/B} + \varphi_b$ applied to the sideband. The switching times are 0.86 ns for the on-to-off transition and 2.70 ns for the off-to-on transition, fully satisfying the requirement of a 10 MHz encoding rate. The implementation of high-speed phase encoding using RF switches is described in Supplementary Note S4.

**Phase compensation and detection.** The optical pulses from Alice and Bob are transmitted to Charlie through 50.4 km of single-mode fibre, respectively. First, channel-induced polarization drift is compensated by a feedback loop consisting of electrically controlled polarization controllers and polarization beam splitters. To suppress the low-frequency, large-amplitude phase drift introduced by long-distance fibre transmission, an auxiliary reference light at a second wavelength $\lambda_2$=1538.48 nm is introduced. This reference light co-propagates with the quantum signal through the same fibre via wavelength-division multiplexing. After demultiplexing at Charlie, the two $\lambda_2$ reference fields are interfered, and the interference signal is amplified by a balanced detector with a gain of approximately $10^6$ before being sent to a PID controller. The error signal generated by the PID controller is fed back to a fibre stretcher (PZT1), achieving coarse locking of the channel phase. The bandwidth of this coarse-locking loop is on the order of kHz, effectively suppressing low-frequency perturbations such as temperature drift and acoustic vibrations.

The optical signals are then interfered at a 50:50 fibre beam splitter. The two output ports are routed through circulators into an FBG, which reflects the carrier and transmits the sideband, thereby spatially separating the two components. In the experiment, to ensure that the carrier and sideband experience the same channel perturbations, they are separated only after interference has taken place. Consequently, the phase feedback effectively acts on subsequent optical pulses, which requires that several successive pulses remain within the coherence time.

The reflected carrier is detected by two gated single-photon detectors (D3 and D4). Their count rates are processed in real time by an FPGA to generate an error voltage proportional to the instantaneous phase difference $\Delta\Phi_{total}(t)$. This signal is fed back through a second fibre stretcher (PZT2), forming a fine-locking feedback loop that dynamically stabilizes the system at the interference maximum, i.e., $\Delta\Phi_{total}(t)\approx 0$. The fine-locking loop has a response time of approximately 20 ms and compensates the residual high-frequency phase noise left after coarse locking.

The transmitted sideband signals are detected by two single-photon detectors (D1 and D2; ID Qube) with a detection efficiency of approximately 10% and a dark count rate of about $2.2\times10^{-7}$ per gate. Charlie records all detection events along with their time tags and publicly announces the valid events through the public channel. Alice and Bob then disclose, for each pulse, the mode and basis they used. Events where both parties selected the encoding mode and Charlie announced a valid detection are retained as the raw key. All remaining events, including those from the decoy mode, are used for parameter estimation and security analysis, from which the final secure key is generated.

# Supplementary Information

## Single-sideband-interference twin-field quantum key distribution without global phase locking

Xingjian Li[1,6†], Bingkun Wang[1,6†], Jianyong Hu[1,2,6†*], Jianqiang Liu[3*], Shuxiao Wu[1,6], Guosheng Feng[4], Zhixing Qiao[4], Changgang Yang[1,6], Ruiyun Chen[1,6], Chengbing Qin[1,6], Guofeng Zhang[1,6], Liantuan Xiao[1,2,5,6*] and Suotang Jia[1,6]

[1]*State Key Laboratory of Quantum Optics Technologies and Devices, Institute of Laser Spectroscopy, Shanxi University, Taiyuan 030006, China*

[2]*Hefei National Laboratory, Hefei 230088, China*

[3]*College of Information Engineering, Shanxi Vocational University of Engineering Science and Technology, Jinzhong 030619, China*

[4]*College of Medical Imaging, Shanxi Medical University, Taiyuan, 030001, China*

[5]*College of Physics and Optoelectronics Engineering, Taiyuan University of Technology, Taiyuan 030024, China*

[6]*Collaborative Innovation Center of Extreme Optics, Shanxi University, Taiyuan 030006, China*

[†]*These authors contribute equally*

**Corresponding author E-mail address: jyhu@sxu.edu.cn; liujianqiang@sxgkd.edu.cn; xlt@sxu.edu.cn*

**Supplementary Note S1: Single Sideband Preparation and Phase Encoding**

We take the optical signal at Alice's end as an example. The initial optical field is $E_A^0(t)=\varepsilon_0\exp\left[i\left(\omega_c t+\theta_A(t)\right)\right]$, where $\varepsilon_0$ is the amplitude of the optical field, $\omega_c$ is the angular frequency of the carrier, and $\theta_A(t)$ is the phase noise of the laser. The IQ modulator consists of two Mach-Zehnder interferometer (MZM) structures, corresponding to the I (in-phase) branch and the Q (quadrature) branch, as shown in Fig. S1. These are modulated by two radio-frequency driving signals, which can be written respectively as:

$$V_I(t)=V_m\cos\left(\Omega t+\varphi_{RF1}\right), V_Q(t)=V_m\cos\left(\Omega t+\varphi_{RF2}\right), \tag{S1}$$

where $\Omega$ is the RF driving signal frequency, $\varphi_{RF1}$ and $\varphi_{RF2}$ are the relative phases of the RF driving signals for the I and Q branches, respectively. $\varphi_{RF1}=\varphi_A^{RF}$ and $\varphi_{RF2}=\varphi_A^{RF}+\Delta\varphi$, where $\varphi_A^{RF}$ is the overall reference phase of Alice's RF source. To simplify the derivation, $\varphi_A^{RF}$ is temporarily omitted in the intermediate steps and will be reintroduced later.

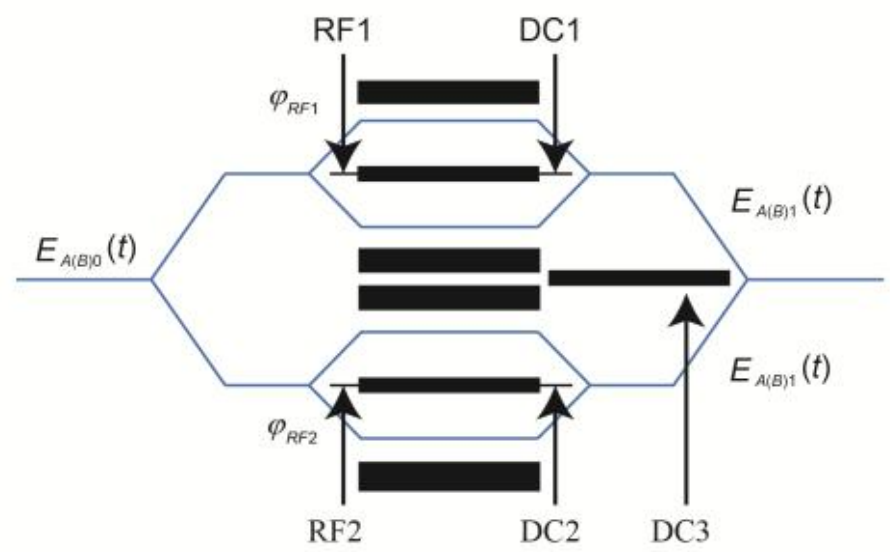

**Figure S1. Internal structure diagram of the IQ modulator.**

The output optical fields of the two arms of the IQ modulator are respectively:

$$E_A^{(1)}(t)=\frac{1}{2}\varepsilon_0 e^{i\left[\omega_c t+\theta_A(t)+\beta\cos(\Omega t+\varphi_{RF1})\right]}, \tag{S2}$$

$$E_A^{(2)}(t)=\frac{1}{2}\varepsilon_0 e^{\left[\omega_c t+\theta_A(t)+\beta\cos(\Omega t+\varphi_{RF2})+\varphi_{DC3}\right]}, \tag{S3}$$

Where $\beta$ is the modulation index and $\varphi_{DC3}$ is the DC bias phase of the IQ modulator. The output optical field of the IQ modulator is:

$$\begin{aligned}E_A(t)&=E_A^{(1)}(t)+E_A^{(2)}(t)\\&=\frac{1}{2}\varepsilon_0 e^{i\left[\omega_c t+\theta_A(t)+\beta\cos(\Omega t+\varphi_{RF1})\right]}+\frac{1}{2}\varepsilon_0 e^{i\left[\omega_c t+\theta_A(t)+\beta\cos(\Omega t+\varphi_{RF2})+\varphi_{DC3}\right]}\\&=\frac{1}{2}\varepsilon_0 e^{i\omega_c t}\left[e^{\left[i\beta\cos(\Omega t+\varphi_{RF1})\right]}+e^{\left[i\beta\cos(\Omega t+\varphi_{RF2})+\right]}+e^{(i\varphi_{DC3})}\right]\end{aligned}. \tag{S4}$$

Using the Bessel function expansion, we obtain:

$$E_A(t)\approx\frac{1}{2}\varepsilon_0 e^{i\left[\omega_c t+\theta_A(t)\right]}\begin{Bmatrix}\left[J_0(\beta)+iJ_1(\beta)e^{i(\Omega t+\varphi_{RF1})}-iJ_{-1}(\beta)e^{-i(\Omega t+\varphi_{RF1})}\right]+\\\left[J_0(\beta)+iJ_1(\beta)e^{i(\Omega t+\varphi_{RF2})}-iJ_{-1}(\beta)e^{-i(\Omega t+\varphi_{RF2})}\right]e^{i\varphi_{DC3}}\end{Bmatrix}. \tag{S5}$$

Substituting $\varphi_{RF1}=\varphi_A^{RF}$ and $\varphi_{RF2}=\varphi_A^{RF}+\Delta\varphi$ into equation (5), we obtain:

$$
\begin{aligned}
E_A(t) &\approx \frac{1}{2}\varepsilon_0 e^{i[\omega_c t+\theta_A(t)]}\left\{\begin{array}{l}\left[J_0(\beta)+iJ_1(\beta)e^{i\left(\Omega t+\varphi_A^{RF}\right)}-iJ_{-1}(\beta)e^{-i\left(\Omega t+\varphi_A^{RF}\right)}\right]+\\ \left[J_0(\beta)+iJ_1(\beta)e^{i\left(\Omega t+\varphi_A^{RF}+\Delta\phi\right)}-iJ_{-1}(\beta)e^{-i\left(\Omega t+\varphi_A^{RF}+\Delta\varphi\right)}\right]e^{i\varphi_{DC3}}\end{array}\right\}\\
&=\frac{1}{2}\varepsilon_0 e^{i[\omega_c t+\theta_A(t)]}\left\{\begin{array}{l} i\left[e^{i\left(\Omega t+\varphi_A^{RF}\right)}\left(1+e^{i(\Delta\varphi+\varphi_{DC3})}\right)\right]J_1(\beta)+\\ \left(1+e^{i\varphi_{DC3}}\right)J_0(\beta)-\\ i\left[e^{-i\left(\Omega t+\varphi_A^{RF}\right)}(1+e^{-i(\Delta\varphi-\varphi_{DC3})})\right]J_{-1}(\beta)\end{array}\right\}
\end{aligned}
\quad . \qquad \text{(S6)}
$$

When $\Delta\varphi$ = -π/2 and $\varphi_{DC3}$ = π/2, the positive first-order sideband is retained.

$$
E_A(t) \approx \frac{1}{2}\varepsilon_0 e^{i[\omega_c t+\theta_A(t)]}\left\{\begin{array}{l} 2e^{i\left(\Omega t+\varphi_A^{RF}+\pi/2\right)}J_1(\beta)+\\ \left(1+e^{i\pi/2}\right)J_0(\beta)\\ 0\end{array}\right\}. \qquad \text{(S7)}
$$

When $\Delta\varphi$ = $\varphi_{DC3}$ = π/2, the negative first-order sideband is retained.

$$
E_A(t) \approx \frac{1}{2}\varepsilon_0 e^{i[\omega_c t+\theta_A(t)]}\left\{\begin{array}{l} 0\\ \left(1+e^{i\pi/2}\right)J_0(\beta)-\\ 2e^{-i(\Omega t+\varphi_A^{RF}-\pi/2)}J_{-1}(\beta)\end{array}\right\}. \qquad \text{(S8)}
$$

Assuming the upper (+1st-order) sideband is generated and considering the initial phase of the light source, the output optical field after the I/Q modulator is obtained as:

$$
\begin{aligned}
E_A(t) &\propto \varepsilon_0\left[J_0(\beta)e^{i[\omega_c t+\theta_A(t)]}+J_1(\beta)e^{i\left[(\omega_c-\Omega)t+\theta_A(t)+\pi/2+\varphi_A^{RF}\right]}\right],\\
&\approx \varepsilon_c e^{i[\omega_c t+\theta_A(t)]}+\varepsilon_s e^{i[(\omega_c-\Omega)t+\theta_A(t)+\varphi_A]}
\end{aligned}
\qquad \text{(S9)}
$$

Where $E_A^C(t)$ is the carrier component in Alice's optical field, and $E_A^S(t)$ is the sideband component; $\varepsilon_c$ and $\varepsilon_s$ are the amplitudes of the carrier and sideband, respectively. A fixed offset of the RF phase $\varphi_A$ is applied, i.e., $\varphi_A^{RF}+\pi/2=\varphi_A$, which is the mathematical manifestation of the strong phase correlation between the carrier and the sideband.

In the encoding mode, Alice maps the bit $b\in\{0,1\}$ to an additional RF signal phase $\varphi_A^{RF}=\varphi_b^A\in\{0,\pi\}$ Therefore, Alice's output optical field is:

$$
E_A(t) \approx \varepsilon_c e^{i[\omega_c t+\theta_A(t)]}+\varepsilon_s e^{i\left[(\omega_c-\Omega)t+\theta_A(t)+\varphi_A+\varphi_b^A\right]}. \qquad \text{(S10)}
$$

The form of Bob's optical field is completely analogous.

$$
E_B(t) \approx \varepsilon_c e^{i[\omega_c t+\theta_B(t)]}+\varepsilon_s e^{i\left[(\omega_c-\Omega)t+\theta_B(t)+\varphi_B+\varphi_b^B\right]}, \qquad \text{(S11)}
$$

Here, we define $\Delta\varphi_b=\varphi_b^A-\varphi_b^B$ as the phase difference encoded by Alice and Bob, which is either 0 or π.

**Supplementary Note S2: Common-Mode Transmission Characteristics of Carrier and Sideband**

After the signals are transmitted through the optical fibre channel to the detection end Charlie, the phase drift introduced is $\delta_A(t)$ or ($\delta_B(t)$). Since the carrier and the sideband are transmitted in the same optical fibre channel and the frequency difference Ω is much smaller than the optical frequency $\omega_c$, the channel disturbances they experience are common-mode. Therefore, the signals emitted by Alice and Bob, upon arriving at Charlie's end, can be written as:

$$E_A^C(t) \approx \begin{Bmatrix} \varepsilon_c e^{i\left[\omega_c t+\theta_A(t-\tau_A)+\delta_A(t)\right]} + \\ \varepsilon_s e^{i\left[(\omega_c-\Omega)t+\theta_A(t-\tau_A)+\delta_A(t)-\varphi_A+\varphi_b^A\right]} \end{Bmatrix}, \tag{S12}$$

$$E_B^C(t) \approx \begin{Bmatrix} \varepsilon_c e^{i\left[\omega_c t+\theta_B(t-\tau_B)+\delta_B(t)\right]} + \\ \varepsilon_s e^{i\left[(\omega_c-\Omega)t+\theta_B(t-\tau_B)+\delta_B(t)-\varphi_B+\varphi_b^B\right]} \end{Bmatrix}, \tag{S13}$$

Where $\tau_{A/B} = L_{A/B} n / c$ is the transmission delay and $n$ is the refractive index of the fibre. At Charlie's end, the carriers from both parties interfere at a 50:50 beam splitter. Consider the optical field at one of the output ports:

$$E_C^+(t) = \frac{1}{\sqrt{2}}\left[E_A^C(t) + E_B^C(t)\right]. \tag{S14}$$

The optical intensity at this port is proportional to $I_C^+(t) \propto \left|E_C^+(t)\right|^2$.

$$\begin{aligned} I_C^+(t) &\propto \Re\left\{E_A^C(t)\left[E_B^C(t)\right]^*\right\} \\ &= \Re\left\{\varepsilon_c\varepsilon_c \exp\left[i\left(\omega_c t + \theta_A(t-\tau_A) + \delta_A(t) - \omega_c t - \theta_B(t-\tau_B) - \delta_B(t)\right)\right]\right\}. \\ &= \varepsilon_c\varepsilon_c \cos\left[\theta_A(t-\tau_A) - \theta_B(t-\tau_B) + \delta_A(t) - \delta_B(t)\right] \end{aligned} \tag{S15}$$

Therefore, by measuring the interference fringes of the carriers, Charlie can extract the total phase difference of the optical signals from Alice and Bob as:

$$\Delta\Phi_{total}(t) = \left[\theta_A(t-\tau_A) - \theta_B(t-\tau_B)\right] + \left[\delta_A(t) - \delta_B(t)\right], \tag{S16}$$

Where $\Delta\varphi_{laser}(t) = \theta_A(t - \tau_A) - \theta_B(t - \tau_B)$ represents the phase difference between the two light sources, and $\Delta\varphi_{channel}(t) = \delta_A(t) - \delta_B(t)$ represents the relative phase drift introduced by the two fibre links. It is noteworthy that the carrier interference is independent of the RF phase $\varphi_A$ and the encoding phase $\varphi_b$. By monitoring $I_{carrier}$ in real time, Charlie can extract the instantaneous total relative phase difference $\Delta\Phi_{total}$ with high precision and use it as a feedback signal for phase compensation. Subsequently, Charlie applies a phase compensation $\Delta\Phi_{total}(t)$ jointly to Bob's carrier and sideband signals after transmission to the detection end. The compensated Bob sideband signal becomes:

$$\begin{aligned} E_{B'}^S(t) &= E_B^S(t)\, e^{i\Delta\Phi_{total}(t)} \\ &= \varepsilon_s e^{i\left[(\omega_c-\Omega)t-\varphi_B+\theta_B(t-\tau_B)+\delta_B(t)+\varphi_b^B+\Delta\Phi_{total}(t)\right]} \end{aligned}. \tag{S17}$$

Now, let the compensated Bob sideband $E_{B'}^S(t)$ interfere with Alice's sideband $E_A^S(t)$. The interference process can be described as:

$$\begin{aligned} I_S^+(t) &\propto \Re\left\{E_A^S(t)\left[E_{B'}^S(t)\right]^*\right\} \\ &= \Re\left\{\varepsilon_s\varepsilon_s \exp\left[i\begin{pmatrix} (\omega_c-\Omega)t - \varphi_A + \theta_A(t-\tau_A) + \delta_A(t) + \varphi_b^A - \\ (\omega_c-\Omega)t + \varphi_B - \theta_B(t-\tau_B) - \delta_B(t) - \varphi_b^B - \Delta\Phi_{total}(t) \end{pmatrix}\right]\right\}. \\ &= \varepsilon_c\varepsilon_c \cos\left[\varphi_B - \varphi_A - \Delta\varphi_b + \Delta\Phi_{total}(t) - \Delta\Phi_{total}(t)\right] \\ &= \varepsilon_c\varepsilon_c \cos\left[\varphi_B - \varphi_A - \Delta\varphi_b\right] \end{aligned} \tag{S18}$$

The final phase difference of the sideband interference is a fixed, time-independent phase, namely the fixed RF

phase difference $\Delta\varphi_{RF} = \varphi_B - \varphi_A = \varphi_B^{RF} - \varphi_A^{RF}$ and the encoding phase difference $\Delta\varphi_b = \varphi_b^B - \varphi_b^A$ that carries the key information. Consequently, the laser and channel noises $\Delta\varphi_{laser}(t)$ and $\Delta\varphi_{channel}(t)$ are effectively suppressed under the premise of system locking. The interference result of the sidebands depends solely on the fixed RF phase difference and the encoding phase. In the experiment, we set the RF phase difference to zero; therefore, the sideband interference result can directly reflect the encoding phases of the two communicating parties.

**Supplementary Note S3: Phase Noise Compensation**

This section elaborates on two key experimental tests. The first is to characterize the residual phase noise of the laser source after locking, to verify that the light source meets the system coherence requirements. The second is to measure the phase noise of the fibre link and demonstrate the effect of a two-level feedback control, consisting of coarse locking based on an auxiliary reference light and fine locking based on the carrier.

(1) Laser Source Phase Noise Test

In this protocol, although global phase locking between the lasers is not required, the coherence of the light sources must still satisfy the fundamental requirement that the coherence time is much longer than the pulse width (1 ns). To this end, we first lock the laser sources of Alice and Bob to a common frequency reference (provided by a laser sent from Charlie's end), and then characterize the residual phase noise of the locked light sources.

The experimental setup is shown in Fig. S2. The two locked lasers are directly input a polarization-maintaining fibre beam splitter for interference, and the interference signal is received and recorded by a photodetector. For comparison, we simultaneously tested the self-interference of a single laser after beam splitting. The results are shown in Fig. S3, where the fringe drift (Fig. S3a) and phase drift rate (Fig. S3b) of the self-interference of a single laser are at a similar level to those of the mutual interference of the two locked lasers (fringe drift in Fig. S3c and phase drift rate in Fig. S3d). Quantitative analysis indicates that the standard deviation of the relative phase drift rate between the two locked lasers is approximately 1.5 times that of a single laser, with the drift characteristic time on the order of seconds. These results demonstrate that, after frequency locking, the relative phase noise of the two independent lasers has been suppressed to an acceptable level, satisfying the interference visibility requirement after transmission over a hundred-kilometer fibre link.

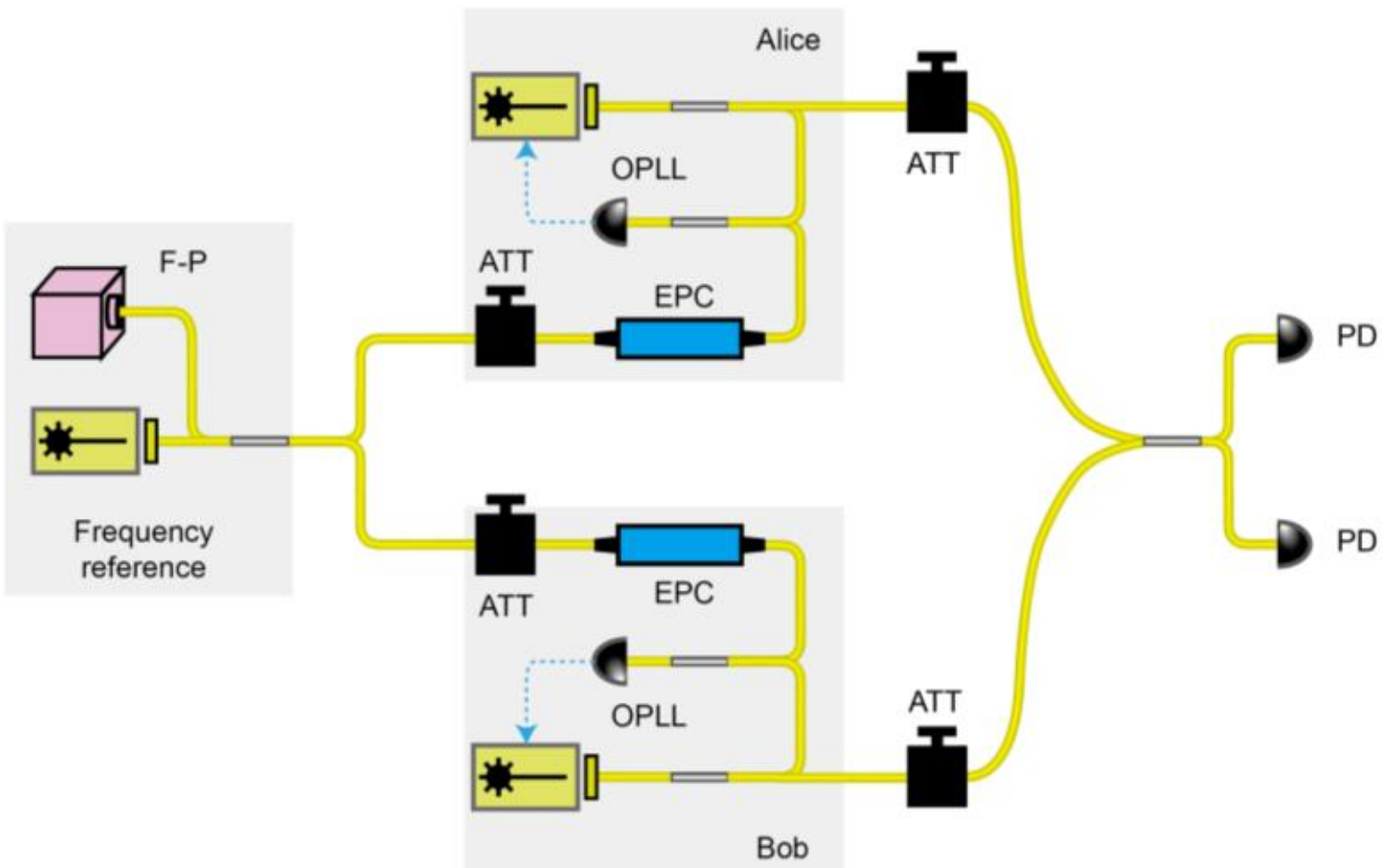

**Figure S2. Experimental setup for laser source noise testing.** The laser sources of Alice and Bob are both locked to a common frequency reference. The output beams from these two light sources directly interfere at a polarization-maintaining beam splitter (BS) and are detected by a photodetector. F-P: Fabry–Perot cavity; OPLL: optical phase-locked loop; ATT: variable optical attenuator; PD:

photodetector; EPC: Electrical control polarization controller.

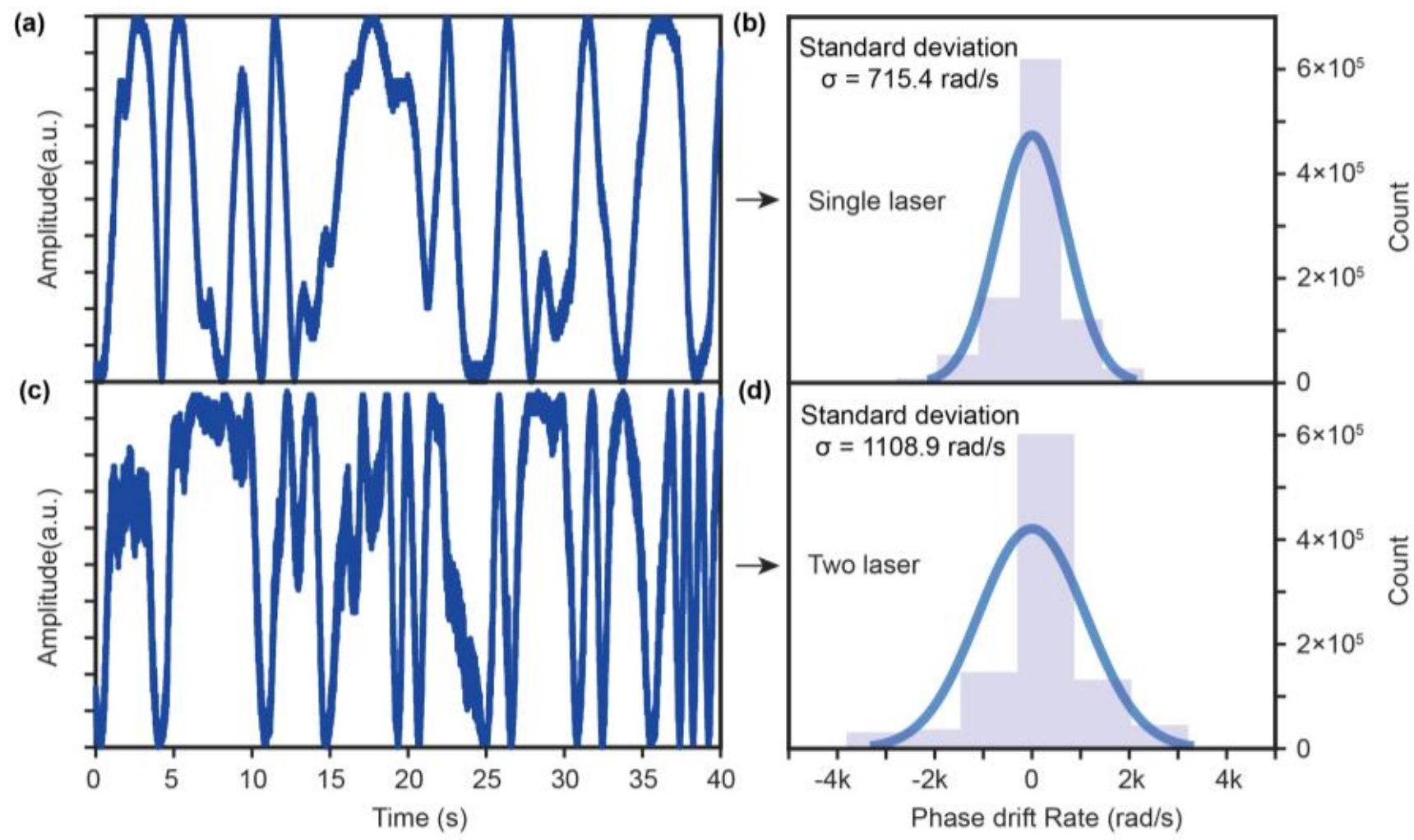

**Figure S3 | Laser source phase noise measurement results. (a)** Interference fringes from self-interference of a single laser. The laser output is split and then interfered with itself to measure its short-term phase stability. The fringes exhibit slow drift on the order of seconds, reflecting the intrinsic phase noise of the laser. **(b)** Phase drift rate from self-interference of a single laser. By tracking the interference fringes in real time, the instantaneous phase variation is calculated. The standard deviation of the phase drift rate is low, indicating good coherence of the single laser. **(c)** Interference fringes from mutual interference of two frequency-locked lasers. The lasers at Alice and Bob are each locked to a common frequency reference (provided by a laser from Charlie), and their outputs are directly interfered. The characteristic drift time of the fringes is similar to that in (a), also on the order of seconds. **(d)** Phase drift rate from mutual interference of two frequency-locked lasers. Quantitative analysis shows that the standard deviation of the relative phase drift rate between the two locked lasers is approximately 1.5 times that of a single laser, remaining within an acceptable range. This result indicates that after frequency locking, the relative phase noise of the two independent lasers is effectively suppressed, meeting the coherence requirement for interference after transmission over a hundred-kilometer fibre link.

(2) Fibre link phase noise and two-level cascaded phase feedback

After confirming that the phase noise of the light source is controllable, we further test and suppress the phase noise introduced by the optical fibre link. The entire link forms an unbalanced Mach-Zehnder interferometer with a total length of 201.6 km (Fig. S4), where phase drift mainly originates from environmental temperature variations and acoustic vibrations. To effectively suppress this link phase noise, we designed a two-level cascaded phase feedback control module.

First level: Coarse locking using an auxiliary laser as frequency reference

To reduce the bandwidth requirement of the phase feedback loop, the system introduces a second wavelength $\lambda_2$ = 1538.48nm as an auxiliary phase reference. This reference light co-propagates with the quantum signal ($\lambda_1$ = 1550.04nm) in the same fibre via wavelength division multiplexing. After demultiplexing at Charlie, interference is performed. The interference signal is monitored by a high-gain balanced detector (gain ~$10^6$), sent to a PID controller to generate an error signal, and then fed back to a fibre stretcher (PZT1) at Alice's end. The bandwidth of this coarse locking loop is on the order of kHz, effectively compensating low-frequency, large-amplitude phase disturbances such as temperature drift and acoustic vibrations.

Figure S5 shows the interference fringe behavior before and after coarse locking is enabled. When the feedback is off, the interference fringes drift rapidly and become difficult to resolve. When the reference light feedback is turned on, the phase drift of both the carrier and the sideband is immediately suppressed to the order of seconds, and the interference fringes become clear and stable. It is worth noting that due to the wavelength difference between $\lambda_1$ and $\lambda_2$, and the fact that the optical paths experienced by the two wavelengths after separation at the transmitting end are not perfectly symmetric, some residual phase noise remains in the $\lambda_1$ signal after coarse locking. This residual noise mainly originates from two sources. First, the initial phase drift difference between the $\lambda_1$ and $\lambda_2$ light sources. Second, the relative phase noise introduced by the unbalanced paths of the two beams after the wavelength division multiplexer at the transmitting end.

Second level: Fine locking using carrier interference

To eliminate the residual phase noise after coarse locking, we further utilize the interference signal of the $\lambda_1$ carrier for fine locking. As shown in Fig. S4, after coarse locking, the $\lambda_1$ signal is interfered, and the carrier is separated using an FBG, then detected by gated single-photon detectors D3 and D4. An FPGA integrates the detector counts over a 20 ms period, extracts an error signal proportional to the instantaneous phase difference $\Delta\Phi_{total}(t)$ via a built-in PID algorithm, and provides real-time feedback to the fibre stretcher PZT2, dynamically locking the system at the interference maximum point (i.e., $\Delta\Phi_{total}(t) \approx 0$).

It should be noted that fibre dispersion effects introduce a fixed relative delay difference between the carrier and the sideband after transmission. In the experiment, this fixed delay is compensated by adjusting the RF phase shifter at the transmitter, ensuring that the residual phase drift of the sideband signal after fine locking is completely suppressed.

Figure S5 shows the result after the two-level feedback control is activated. Under the combined action of coarse and fine locking, the interference fringes of both the carrier and the sideband stabilize to the Hz level, and the average interference visibility over a long period (more than one hour) remains above 96.9%, laying a solid foundation for highly stable quantum key generation.

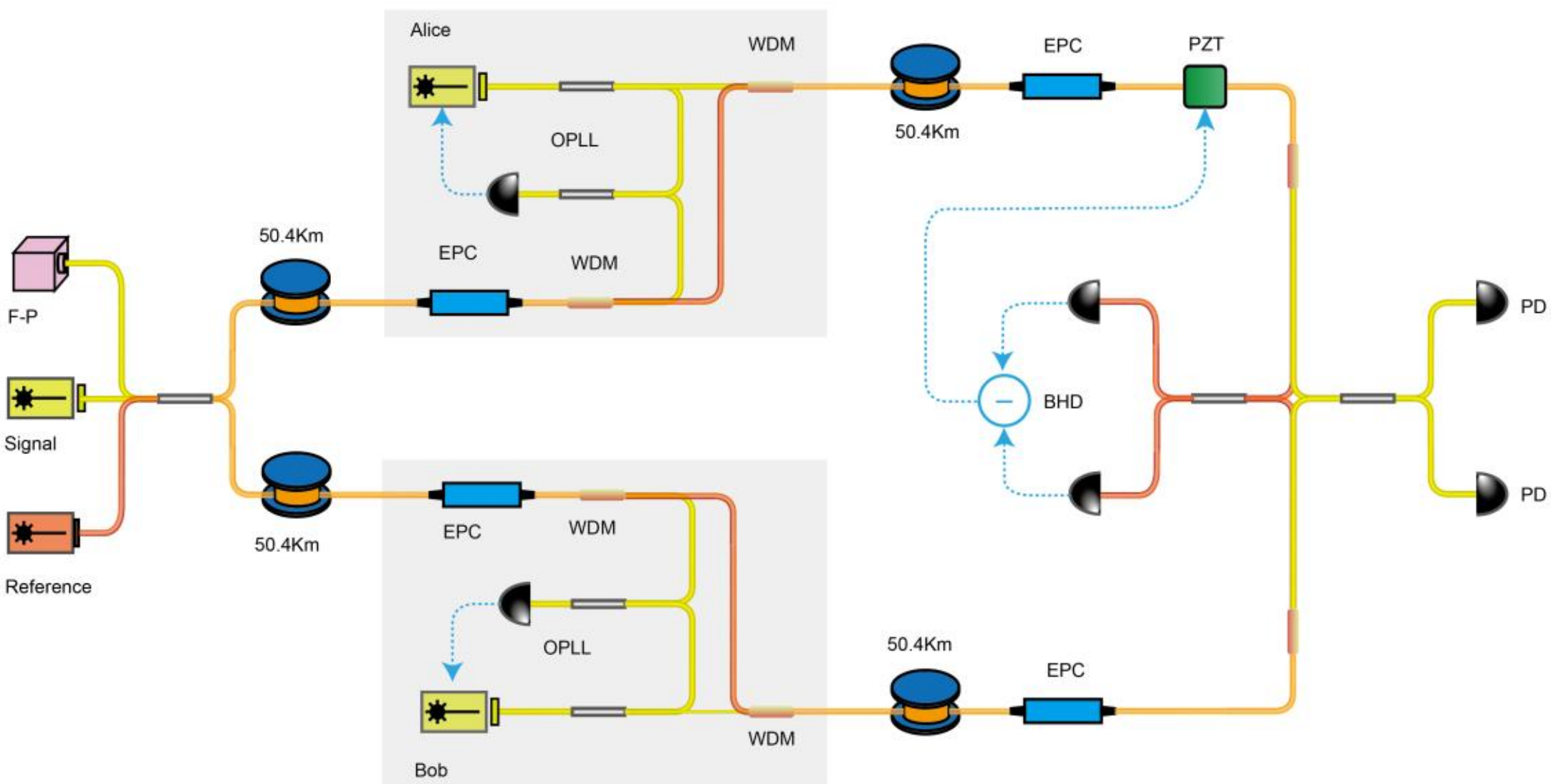

**Figure S4. Experimental setup for fibre link phase noise and strong reference signal feedback testing.** The entire test setup can be viewed as a 201.6 km unbalanced Mach-Zehnder interferometer. In this figure, the red line represents the optical path of the reference light, the yellow line represents the optical path of the signal light, the blue dashed line represents the electrical signal circuit, and the orange fiber represents the optical path after wavelength division multiplexing of the reference light and the signal light. BHD: balanced detector.

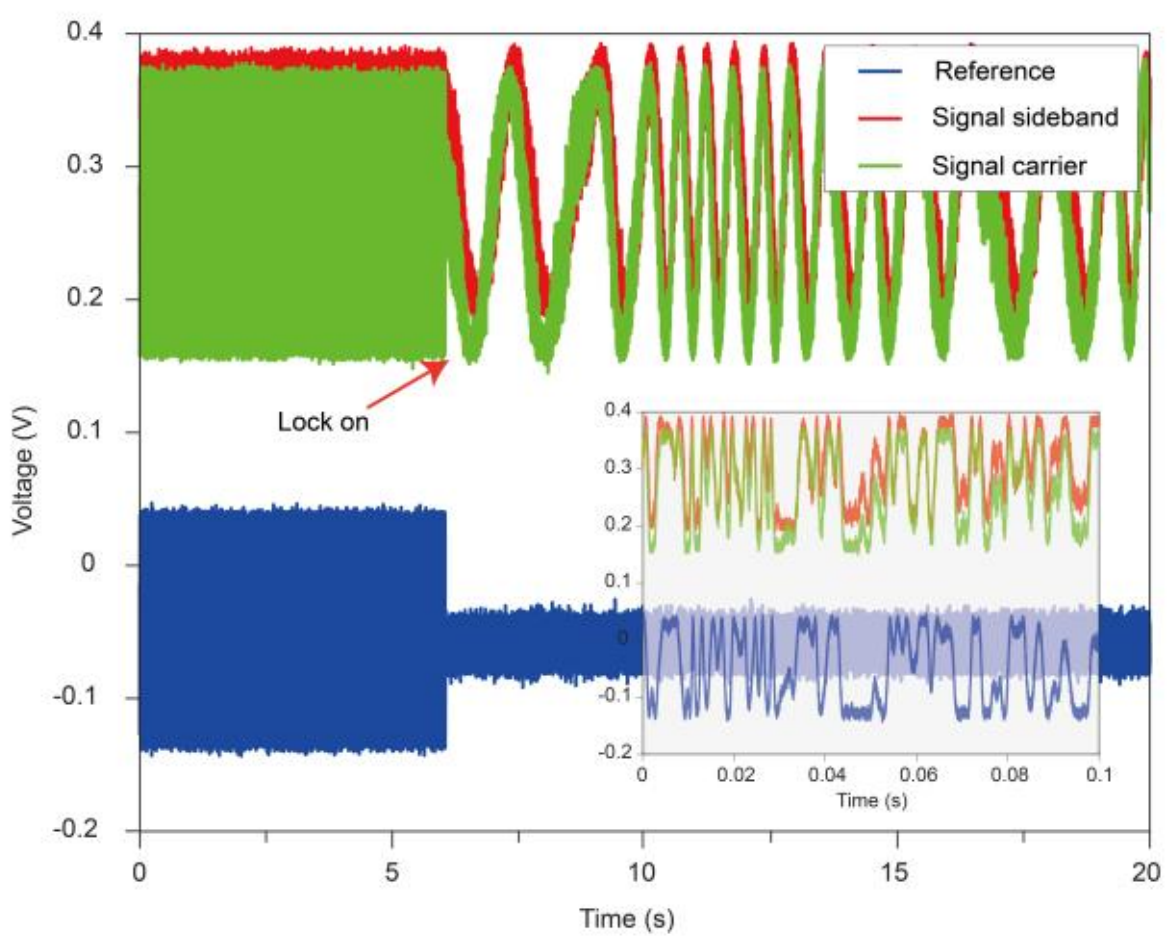

**Figure S5. Test results of fibre link phase noise.** In the figure, blue represents the reference light interference signal, red represents the sideband interference signal, and green represents the carrier interference signal. The inset shows the detailed interference fringes within the 0-0.1 second time window. After the reference light feedback is activated, the phase drift is significantly suppressed to the order of seconds.

## Supplementary Note S4: Phase Encoding implementation

This section elaborates on the implementation of phase encoding in the SSB-TF-QKD system. As described in the main text, the key information is encoded by manipulating the phase of the sideband, which is ultimately determined by the phase of the RF driving signal. Therefore, the core of phase encoding lies in the high-speed random switching of the RF signal phase.

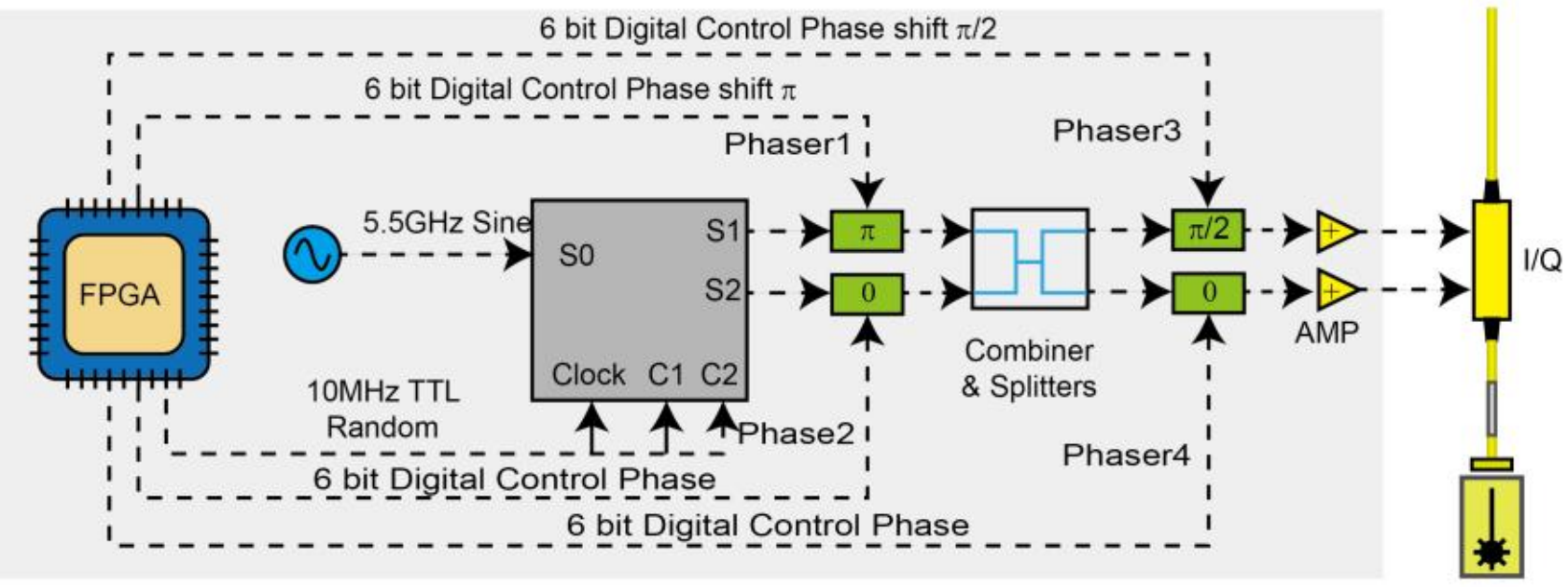

**Figure S6. Schematic diagram of phase encoding based on sideband interference.** Field-Programmable Gate Array (FPGA); Numerically Controlled RF Phase Shifter (Phaser); RF Amplifier (AMP).

Figure S6 illustrates the phase encoding structure based on sideband interference. The entire encoding system is synchronized by a global clock reference provided by a digital delay pulse generator at Charlie's end. After electro-optical conversion, this reference is distributed via a synchronization fibre to the FPGAs at both Alice and Bob, ensuring strict alignment of pulse timing between the two parties.

At Alice's end (Bob's end has an identical structure), the FPGA generates a preset pseudo-random number sequence in real time at a repetition rate of 10 MHz. This random number directly controls the channel selection of a high-speed RF switch. The two output ports of the RF switch are connected to two numerically controlled phase shifters (Phaser1 and Phaser2), which are preset to different phase values. One corresponds to an encoding phase of 0, and the other corresponds to an encoding phase of π. When the RF switch selects the path connected to Phaser1, the final encoding phase imparted to the sideband is 0; when it selects the path connected to Phaser2, the encoding phase is π.

A 5.5 GHz RF signal is used to generate the sideband. Its output continuous sine wave is split into two paths by a power divider. These two paths respectively enter the two aforementioned numerically controlled phase shifters. After passing through the phase shifters, the two signals are combined by a power combiner. The combined signal is then split again by a second-stage power divider into two signals of equal amplitude and precisely correlated phase, which serve as the I and Q drive inputs for the IQ modulator. Before being fed into the IQ modulator, each of these two driving signals passes through an additional stage of numerically controlled phase shifters for fine adjustment. This ensures that the phase difference between the I and Q paths is precisely locked at $\pi/2$, thereby achieving single-sideband modulation. Finally, the two signals are amplified to the required power by RF amplifiers before being applied to the IQ modulator.

It is important to emphasize that the design of the entire RF link ensures the stability of the encoding phase: the RF switch selects between pre-calibrated, fixed phase shifter paths, rather than performing continuous phase modulation on the RF signal itself. This avoids the phase noise and instability that active phase shifting may introduce.

To verify whether the phase encoding system meets the requirement of a 10 MHz encoding rate, the switching time of the RF switch was measured. The test results are shown in Fig. S7. Figures S7(a) and (b) show the switching response speed test for the path containing the numerically controlled phase shifter Phaser1. Figure S7(a) displays the timing relationship between the control signal output by the FPGA (yellow curve) and the output signal of the RF switch (blue curve), while Fig. S7(b) shows a magnified view of the switching edge. The measurement indicates that the switch's response time from on to off is approximately 0.86 ns. Figures S7(c) and (d) show the corresponding test results for the path containing Phaser2. Figure S7(c) displays the timing relationship between the control signal and the output signal, and Fig. S7(d) shows a magnified view of the switching edge. The measurement indicates that the switch's response time from off to on is approximately 2.70 ns. These switching times are significantly shorter than the 10 MHz encoding period (100 ns), demonstrating that the phase encoding system has ample timing margin and can stably support a 10 MHz encoding rate, meeting the design requirements of the experimental system.

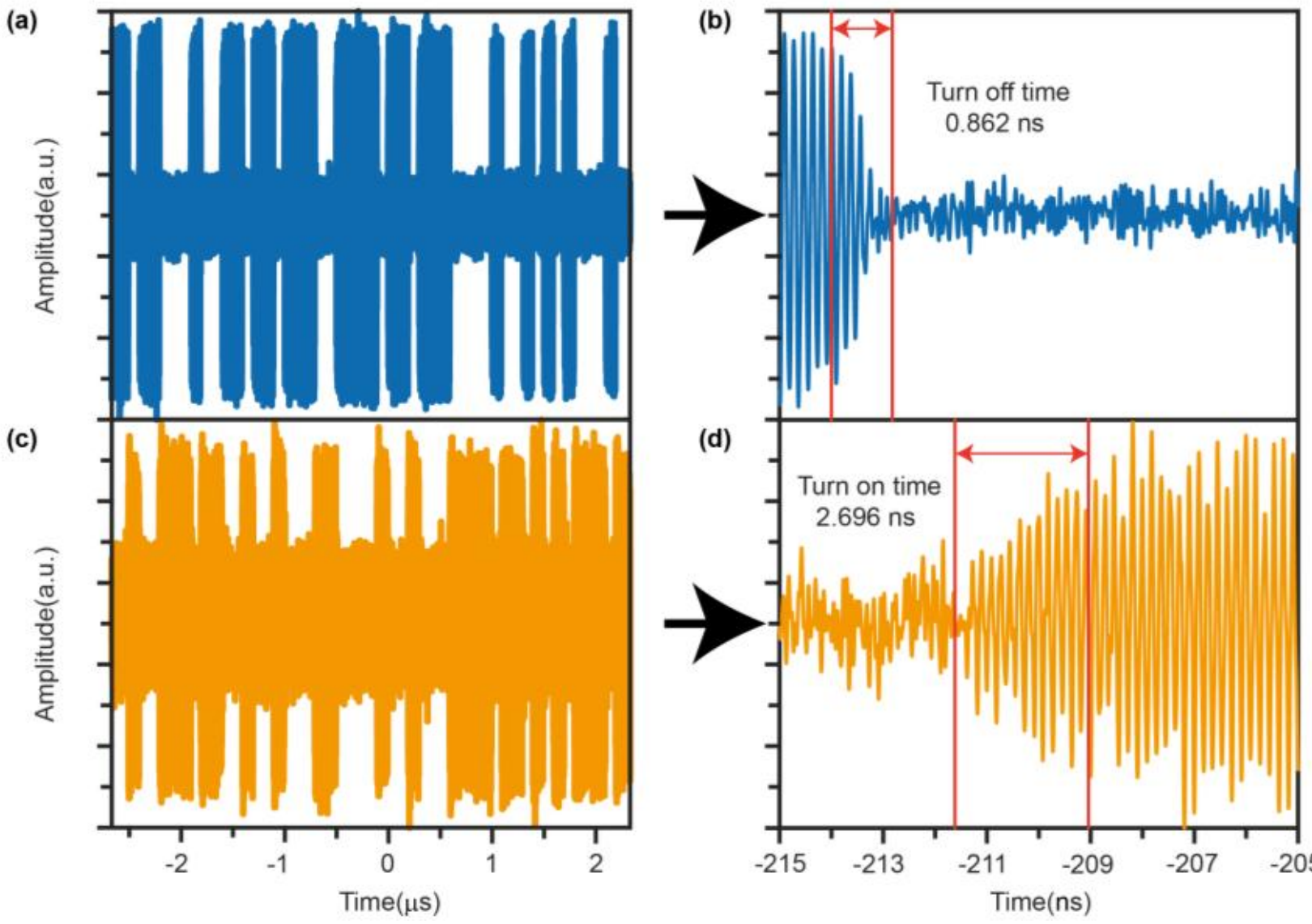

**Figure S7. Test results of the phase encoding system switching time.** (a) Switching response test for the Phaser1 path. (b) Measurement of the turn-off time for the Phaser1 path. The falling edge of the switch transitioning from on to off in (a) is magnified, and the measured response time is 0.86 ns. This time corresponds to the duration required for the RF output amplitude to drop from 90% to 10%. (c) Switching response test for the Phaser2 path. (d) Measurement of the turn-on time for the Phaser2 path. The rising edge of the switch

transitioning from off to on in (c) is magnified, and the measured response time is 2.70 ns. This time corresponds to the duration required for the RF output amplitude to rise from 10% to 90%.

**Supplementary Note S5: Security Analysis and Secure Key Rate Calculation**

This section elaborates on the security proof framework for the SSB-TF-QKD protocol and provides the specific formula for calculating the secure key rate. The analysis shows that although the protocol introduces a strong carrier as a phase reference, since the carrier carries no key information and its measurement results can be transmitted via a public channel, the protocol can still be rigorously analyzed for security within the standard measurement-device-independent (MDI) security framework [1-3].

**S5.1 Security Analysis Framework**

This protocol can be incorporated into the security framework of measurement-device-independent quantum key distribution (MDI-QKD). Charlie, as an untrusted third party, can announce all his measurement results (including the phase difference $\Delta\Phi_{total}(t)$ extracted from carrier interference) via a public channel. Although the carrier is a strong classical light, it carries no key information—the key is encoded only in the phase of the sideband (0 or π) and is independent of the carrier. Therefore, publicly disclosing the carrier phase information does not lead to the leakage of key information.

The phase feedback signal is "common mode." The publicly disclosed phase error signal $\Delta\Phi_{total}(t)=\Delta\varphi_{laser}(t)+\Delta\varphi_{channel}(t)$ reflects the overall noise characteristics of the channel and lasers. This is a common reference signal and is transparent to the encoding of Alice and Bob. Eve's acquisition of this signal does not help her distinguish whether a specific detection event was caused by a relative phase difference of 0 or π between Alice and Bob. This is analogous to the public global phase reference in phase-encoding MDI-QKD, the security of which has been rigorously proven.

The decoy-state method is employed to against photon-number-splitting attacks. Alice and Bob randomly send phase-randomized coherent states with average photon numbers of $|\alpha|^2$ (signal state), $\mu$, $\nu$, and $\omega$ (decoy states, satisfying $\omega \ll \nu < \mu$). By comparing the gains and error rates under different intensity combinations, the contribution of the single-photon components can be rigorously estimated, thereby allowing for the calculation of the secure key rate.

**S5.2 Secure Key Rate Formula**

The secure key rate formula for TF-QKD based on the CAL (Curty-Azuma-Lo) protocol[1–3] is:

$$R \geq \max\{R_{10}, 0\} + \max\{R_{01}, 0\}, \tag{S19}$$

Where $R_{D_0D_1}$ represents the secure key rate under the condition that the two detectors for sideband detection respond accordingly, with $(D_0, D_1) \in \{(1,0), (0,1)\}$. According to GLLP, $R_{D_0D_1}$ can be defined as:

$$R_{D_0,D_1} = q \cdot Q^X_{\alpha\alpha,D_0D_1}\left[1 - H_2\left(e^X_{bit,D_0D_1}\right) - H_2\left(e^U_{ph,D_0D_1}\right)\right]. \tag{S20}$$

Here, $H_2(x) = -x\log_2 x - (1 - x)\log_2(1 - x)$ is the binary entropy function. $q$ denotes the probability that both Alice and Bob both select the encoding mode. $Q^X_{\alpha\alpha,D_0D_1}$ represents the total gain in the encoding mode (with an average photon number of $|\alpha|^2$). It is the probability that Charlie announces outcomes (0,1) and (1,0) given that Alice and Bob are both in the encoding mode. $e^X_{bit,D_0D_1}$ is the bit error rate for the corresponding events, indicating the proportion of mismatched key bits after sifting under the condition that both parties have selected the encoding mode. $e^{U}_{ph,D_0D_1}$ represents the phase error rate, which cannot be directly measured and must be estimated via an upper bound using decoy-state data. The following sections will elaborate on the specific definitions and calculation methods for each of these parameters.

**S5.2.1 The Total Gain $Q^{X}_{\alpha\alpha,D_0D_1}$**

The total gain in the encoding mode is defined as: the probability that Charlie announces a valid detection event (i.e., exactly one of the detectors $D_0$ and $D_1$ clicks) given that both Alice and Bob have selected the encoding mode and both have sent signal states with an average photon number $|\alpha|^2$. This gain can be directly obtained through statistical analysis of the experimental data:

$$\begin{aligned} Q^{X}_{\alpha\alpha,D_0D_1} &= \sum\nolimits_{b_A,b_B\in\{0,1\}} p\left(b_A,b_B\right) p_{click}\left(D_0,D_1\,|\,b_A,b_B\right) \\ &= \frac{1}{4}\sum\nolimits_{b_A,b_B\in\{0,1\}} p_{click}\left(D_0,D_1\,|\,b_A,b_B\right) \end{aligned}, \tag{S21}$$

where $p(b_A, b_B)$ represents the joint probability that Alice and Bob prepare signal states corresponding to the bit pair $(b_A, b_B)$. In this protocol, each bit pair is chosen uniformly at random, so $p(b_A, b_B)=1/4$ (holds for all $b_A$, $b_B$). $p_{click}(D_0, D_1|b_A, b_B)$ denotes the conditional probability that Charlie announces the detection result $(b_A, b_B)$, given that Alice and Bob transmit signal states corresponding to $b_A$ and $b_B$, respectively. Here, $(D_0, D_1)\in\{(1,0), (0,1)\}$ correspond to a click event. The above conditional probability can be directly estimated by statistically analyzing the detection counts from the experimental data in the encoding mode.

**S5.2.2 Calculation of the Bit Error Rate $e^{X}_{bit,D_0D_1}$**

The bit error rate is defined as: the proportion of mismatched raw key bits held by Alice and Bob among the sifted valid events. According to the physical principle of sideband interference, when the encoding phases of both parties are identical ($\Delta\varphi_b=0$), photons tend to be detected by $D_0$; when the encoding phases are opposite ($\Delta\varphi_b=\pi$), photons tend to be detected by $D_1$. Therefore, the formula for the bit error rate can be expressed as:

$$e^{X}_{bit,10} = \frac{\sum_{i=1}^{1} p\left(D_0=1,D_1=0\,|\,b_A=i,b_B=i\right)}{4Q^{X}_{\alpha\alpha,10}}, \tag{S22}$$

$$e^{X}_{bit,01} = \frac{\sum_{i,j|i\oplus j=1}^{1} p\left(D_0=0,D_1=1\,|\,b_A=i,b_B=j\right)}{4Q^{X}_{\alpha\alpha,01}}. \tag{S23}$$

**S5.2.3 Estimation of the Upper Bound of the Phase Error Rate $e^{U}_{ph,D_0D_1}$**

The phase error rate $e^{U}_{ph,D_0D_1}$ corresponds to the bit information that Eve can obtain. It cannot be directly measured and must be estimated from the decoy-state method. The core idea is to use gain equations under different intensity combinations to extract the contribution of the single-photon components, thereby obtaining an upper bound for the phase error rate.

(1) Gain Equations and the Decoy-State Method

The phase error rate $e^{U}_{ph,D_0D_1}$ is given by the following formula:

$$e^{U}_{ph,D_0D_1} = \frac{1}{Q^{X}_{\alpha\alpha,D_0D_1}}\left[\left(\sum\nolimits_{n,m=0}^{\infty} c_{2n}c_{2m}\sqrt{Y_{2n\,2m\,D_0D_1}}\right)^2 + \left(\sum\nolimits_{n,m=0}^{\infty} c_{2n+1}c_{2m+1}\sqrt{Y_{2n+1\,2m+1\,D_0D_1}}\right)^2\right], \tag{S24}$$

Where $c_n = e^{-|\alpha|^2/2}\alpha^n/\sqrt{n!}$ ($n$ and $m$ represent the photon numbers, respectively). $Y_{2n\,2m\,D_0D_1}$ denotes the conditional probability that Charlie obtains the detection result given that Alice and Bob send $n$-photon and $m$-photon states, respectively. Physically, the first term describes the interference component when Alice and Bob each emit even-photon-number states, while the second term corresponds to the interference component for odd-photon-number states.

(2) Expression for the Upper Bound of the Phase Error Rate $e^{U}_{ph,\,D_0D_1}$

Since we can only use a finite number of decoy-state intensities, it is impossible to solve for all $Y_{n,m}$ individually. Instead, we can only obtain an upper bound for them, thereby estimating the maximum possible value of $e^{U}_{ph,\,D_0D_1}$ for secure key rate calculation

$$e^{U}_{ph,\,D_0D_1} \le \frac{1}{Q^{X}_{\alpha\alpha,D_0D_1}}\left[\begin{array}{l}\left(\sum_{(2n,2m)\in H}^{\infty} c_{2n}c_{2m}\sqrt{Y^{U}_{2n.2m\,D_0,D_1}} + \sum_{(2n,2m)\notin H}^{\infty} c_{2n}c_{2m}\right)^2 \\ +\left(\sum_{(2n+1,2m+1)\in H}^{\infty} c_{2n+1}c_{2m+1}\sqrt{Y^{U}_{2n+1.2m+1\,D_0,D_1}} + \sum_{(2n+1,2m+1)\notin H}^{\infty} c_{2n+1}c_{2m+1}\right)^2\end{array}\right], \tag{S25}$$

where $Y_{2n\,2m\,D_0D_1}$ is a conditional probability, which necessarily satisfies $0 \le Y_{2n2m,D_0D_1} \le 1$; therefore, its upper bound is taken as 1. Thus, in Eq.(S25), the terms for $(n,m)\notin H$ are represented by their extreme contribution using the product of coherent state coefficients $c_nc_m$. Furthermore, since only terms with an even total photon number contribute to the phase error rate expression, the summations above are restricted to $(2n,2m)$ and $(2n+1, 2m+1)$, i.e., only even-even and odd-odd photon number pairs are included. The notation $Y^{U}_{nm,D_0D_1}$ represents the upper bound estimate of $Y_{nm\,D_0D_1}$ for a given detection outcome ($D_0$, $D_1$), while $Q^{X}_{\alpha\alpha,D_0D_1}$ s the total probability that Charlie announces exactly one click, i.e., $(D_0, D_1)\in\{(1, 0), (0, 1)\}$.

In the decoy-state method, Alice and Bob each emit coherent state pulses with intensities $\mu_a$ and $\mu_b$, where $(\mu_a, \mu_b)\in\{\mu, v, \omega\}$, and Charlie announces the detection outcomes $(D_0, D_1)\in\{(1, 0), (0, 1)\}$. The total gain (gain) that can be directly measured experimentally is defined as:

$$Q^{(D_0,D_1)}_{\mu_a,\mu_b} = e^{-(\mu_a+\mu_b)}\sum_{n,m}^{\infty}\frac{\mu_a^n\mu_b^m}{n!m!}Y_{nm,D_0D_1}\ , \tag{S26}$$

where $Y^{U}_{nm,D_0D_1}$ represents the conditional probability that Charlie announces the outcome ($D_0$, $D_1$) given that Alice sends an $n$-photon state and Bob sends an $m$-photon state. By selecting different intensity combinations in the experiment, such as $\{\mu\mu, v\mu, \mu v, \omega\mu, \mu\omega, v\omega, \omega v, vv, \omega\omega\}$, the corresponding gain equations can be obtained, thereby enabling the derivation of the upper bounds for the respective yields.

(3) Estimation Formulas for Key Yields

Calculation of $Y^{U}_{00}$

$$
\begin{aligned}
\Delta_{\mu\nu} &= \nu^2 e^{2\mu} Q_{\mu\mu}^{D_0,D_1} + \mu^2 e^{2\nu} Q_{\nu\nu}^{D_0,D_1} - \mu\nu e^{\mu+\nu}\left(Q_{\mu\nu}^{D_0,D_1} + Q_{\nu\mu}^{D_0,D_1}\right) \\
\Delta_{\mu\omega} &= \omega^2 e^{2\mu} Q_{\mu\mu}^{D_0,D_1} + \mu^2 e^{2\omega} Q_{\omega\omega}^{D_0,D_1} - \mu\omega e^{\mu+\omega}\left(Q_{\mu\omega}^{D_0,D_1} + Q_{\omega\mu}^{D_0,D_1}\right). \\
\Delta_{\nu\omega} &= \omega^2 e^{2\nu} Q_{\mu\mu}^{D_0,D_1} + \nu^2 e^{2\omega} Q_{\omega\omega}^{D_0,D_1} - \nu\omega e^{\nu+\omega}\left(Q_{\nu\omega}^{D_0,D_1} + Q_{\omega\mu}^{D_0,D_1}\right)
\end{aligned}
\tag{S27}
$$

Then the upper bound for the zero-photon yield $Y_{00}^{U}$ is given by:

$$
Y_{00}^{U} = \frac{\dfrac{\omega^2\Delta_{\mu\nu}}{\mu-\nu} - \dfrac{\nu^2\Delta_{\mu\omega}}{\mu-\omega} + \dfrac{\mu^2\Delta_{\nu\omega}}{\nu-\omega}}{(\mu-\nu)(\mu-\omega)(\nu-\omega)}. \tag{S28}
$$

Calculation of $Y_{11}^{U}$

$$
T_{\mu\nu} = e^{\mu+\nu} Q_{\nu\nu}^{D_0,D_1} + e^{\mu+\mu} Q_{\mu\mu}^{D_0,D_1} - e^{\mu+\nu}\left(Q_{\mu\nu}^{D_0,D_1} + Q_{\nu\mu}^{D_0,D_1}\right). \tag{S29}
$$

Its preliminary upper bound is: $Y_{11}^{U'} = \dfrac{T_{\mu\nu}}{(\mu-\nu)^2}$

Considering higher-order correction terms, define:

$$
\begin{aligned}
\xi_1 &= \left(\nu^2-\omega^2\right)^2\left[e^{\mu+\mu} Q_{\mu\mu}^{D_0,D_1} - e^{\mu+\omega} Q_{\mu\omega}^{D_0,D_1} - e^{\omega+\mu} Q_{\omega\mu}^{D_0,D_1} + e^{\omega+\omega} Q_{\omega\omega}^{D_0,D_1}\right] \\
&\quad - \left(\nu^2-\omega^2\right)\left(\mu^2-\omega^2\right)\left[\left(\nu^2-\omega^2\right) - e^{\mu+\omega} Q_{\mu\omega}^{D_0,D_1} - e^{\omega+\nu} Q_{\omega\nu}^{D_0,D_1} + e^{\omega+\omega} Q_{\omega\omega}^{D_0,D_1}\right] \\
&\quad - \left(\mu^2-\omega^2\right)\left(\nu^2-\omega^2\right)\left[e^{\nu+\mu} Q_{\nu\omega}^{D_0,D_1} - e^{\nu+\omega} Q_{\nu\omega}^{D_0,D_1} - e^{\omega+\mu} Q_{\omega\mu}^{D_0,D_1} + e^{\omega+\omega} Q_{\omega\omega}^{D_0,D_1}\right]. \\
&\quad + \left(\mu^2-\omega^2\right)^2\left[e^{\nu+\nu} Q_{\nu\nu}^{D_0,D_1} - e^{\nu+\omega} Q_{\nu\omega}^{D_0,D_1} - e^{\omega+\mu} Q_{\omega\nu}^{D_0,D_1} + e^{\omega+\omega} Q_{\omega\omega}^{D_0,D_1}\right] \\
\xi_2 &= \left[\left(\nu^2-\omega^2\right)(\mu-\omega) - \left(\mu^2-\omega^2\right)(\nu-\omega)\right]^2
\end{aligned}
\tag{S30}
$$

Then: $Y_{11}^{U''} = \dfrac{\xi_1}{\xi_2}$

$$
Y_{11}^{U} = \min\left(Y_{11}^{U'}, Y_{11}^{U''}\right). \tag{S31}
$$

Calculation of $Y_{02}^{U}$

Define:

$$
\begin{aligned}
\Gamma &= (\mu-\nu)\left(\mu e^{\omega+\omega} Q_{\omega\omega}^{D_0,D_1} - \omega e^{\mu+\omega} Q_{\mu\omega}^{D_0,D_1}\right) \\
&\quad + (\mu-\omega)\left(\omega e^{\mu+\nu} Q_{\mu\nu}^{D_0,D_1} - \mu e^{\omega+\nu} Q_{\omega\nu}^{D_0,D_1}\right) \\
&\quad + (\nu-\omega)\left(\mu e^{\omega+\nu} Q_{\omega\mu}^{D_0,D_1} - \omega e^{\mu+\mu} Q_{\mu\mu}^{D_0,D_1}\right) \\
\Lambda &= \left[\omega\left(1-e^{\mu}\right) - \mu\left(1-e^{\omega}\right)\right]\begin{bmatrix}(\nu-\omega)\left(\omega-\mu-e^{\omega}+e^{\mu}\right) - \\ (\mu-\omega)\left(\mu-\nu-e^{\omega}+e^{\nu}\right)\end{bmatrix}
\end{aligned}
\tag{S32}
$$

Then: $Y_{02}^{U} = \dfrac{2(\Gamma-\Lambda)}{(\mu-\omega)^2(\nu-\omega)(\mu-\nu)}$

Calculation of $Y_{20}^{U}$

Define:

$$\begin{aligned}
\Xi &= (\mu-\nu)\left(\nu e^{\omega+\omega} Q_{\omega\omega}^{D_0,D_1} - \omega e^{\nu+\omega} Q_{\nu\omega}^{D_0,D_1}\right) \\
&+(\nu-\omega)\left(\omega e^{\nu+\mu} Q_{\nu\mu}^{D_0,D_1} - \nu e^{\omega+\mu} Q_{\omega\mu}^{D_0,D_1}\right) \\
&+(\mu-\omega)\left(\nu e^{\omega+\nu} Q_{\omega\nu}^{D_0,D_1} - \omega e^{\nu+\nu} Q_{\nu\nu}^{D_0,D_1}\right) \\
\Upsilon &= \left[\omega\left(1-e^{\nu}\right)-\nu\left(1-e^{\omega}\right)\right]\begin{bmatrix}(\mu-\omega)\left(\omega-\nu-e^{\omega}+e^{\nu}\right)- \\ (\nu-\omega)\left(\omega-\mu-e^{\omega}+e^{\mu}\right)\end{bmatrix}
\end{aligned} \qquad \text{(S33)}$$

Then: $Y_{20}^{U} = \dfrac{2(\Xi-\Upsilon)}{(\nu-\omega)^2(\mu-\omega)(\mu-\nu)}$

**Supplementary Note S6: Detailed Experimental Data**

To calculate the secure key rate corresponding to each channel loss condition, Table S1 presents the values of parameters $Q_{\alpha\alpha,D_0D_1}^{X}$, $e_{bit,D_0D_1}^{X}$, $Y_{nm,D_0D_1}^{U}$, $e_{ph,D_0D_1}^{U}$, and $R_{D_0D_1}$ for three loss settings shown in Figure 3, namely 33.7 dB, 42.9 dB, and 52.9 dB. Here ($D_0$, $D_1$)∈{(1, 0), (0, 1)} and ($n$, $m$)∈{(0, 0), (0, 2), (2, 0), (1, 1)}. for three different values of total system loss: 33.7 dB, 42.9 dB, and 52.9 dB. As described in the main text, for the 33.7 dB loss, the service link length is 100.8 km and the quantum link length is 0 km; for 42.9 dB, links are 100.8 km; for 52.9 dB, an additional 10 dB attenuation is added to the quantum link. Parameters $Q_{\alpha\alpha,D_0D_1}^{X}$ and $e_{bit,D_0D_1}^{X}$ were directly obtained from experimental data corresponding to signals in the encoding mode using Eq. (S21)-(S23). Using Eq. (S24)-(S32), we calculated the yields $Y_{nm,D_0D_1}^{U}$ for ($n$, $m$)∈{(0, 0), (0, 2), (2, 0), (1, 1)}, together with the experimentally observed gains $Q_{\mu_a,\mu_b}^{D_0,D_1}$ provided in Table S2, where ($\mu_a$, $\mu_b$)∈{$\mu$, $\nu$, $\omega$}correspond to the decoy state intensity. Finally, the phase error rate $e_{ph,D_0D_1}^{U}$ was determined, and the secret key rate $R_{D_0D_1}$ was estimated using Eq. (S19)-(S20).

**Table S1. Values of parameters**

| Loss : 33.7 dB | $e_{bit,D_0D_1}^{X}$ | $Q_{\alpha\alpha,D_0D_1}^{X}$ | $Y_{00,D_0D_1}^{U}$ | $Y_{11,D_0D_1}^{U}$ | $Y_{02,D_0D_1}^{U}$ | $Y_{20,D_0D_1}^{U}$ | $e_{ph,D_0D_1}^{U}$ | $R_{D_0D_1}$ |
|---|---|---|---|---|---|---|---|---|
| $D_0=1$, $D_1=0$ | 2.0408e-2 | 3.1756e-4 | 2.8278e-7 | 4.2685e-3 | 4.9111e-3 | 2.9679e-3 | 1.9497e-3 | 6.4537e-5 |
| $D_0=0$, $D_1=1$ | 1.8518e-2 | 3.8028e-4 | 2.1924e-7 | 2.7141e-3 | 2.5969e-3 | 2.8306e-3 | 1.4222e-3 | 7.8924e-5 |
| Loss : 33.7 dB | $e_{bit,D_0D_1}^{X}$ | $Q_{\alpha\alpha,D_0D_1}^{X}$ | $Y_{00,D_0D_1}^{U}$ | $Y_{11,D_0D_1}^{U}$ | $Y_{02,D_0D_1}^{U}$ | $Y_{20,D_0D_1}^{U}$ | $e_{ph,D_0D_1}^{U}$ | $R_{D_0D_1}$ |
| $D_0=1$, $D_1=0$ | 2.0408e-2 | 3.1756e-4 | 4.5862e-7 | 2.0516e-3 | 3.3478e-3 | 3.0329e-3 | 2.5468e-3 | 6.4120e-5 |
| $D_0=0$, $D_1=1$ | 1.8518e-2 | 3.8028e-4 | 2.6445e-7 | 5.5291e-3 | 4.2566e-3 | 1.0591e-3 | 1.4393e-3 | 7.8951e-5 |
| Loss : 33.7 dB | $e_{bit,D_0D_1}^{X}$ | $Q_{\alpha\alpha,D_0D_1}^{X}$ | $Y_{00,D_0D_1}^{U}$ | $Y_{11,D_0D_1}^{U}$ | $Y_{02,D_0D_1}^{U}$ | $Y_{20,D_0D_1}^{U}$ | $e_{ph,D_0D_1}^{U}$ | $R_{D_0D_1}$ |
| $D_0=1$, $D_1=0$ | 2.0408e-2 | 3.1756e-4 | 2.0353e-7 | 3.4536e-3 | 3.7801e-3 | 1.6135e-3 | 1.3862e-3 | 6.4950e-5 |
| $D_0=0$, $D_1=1$ | 1.8518e-2 | 3.8028e-4 | 1.1108e-7 | 3.1723e-3 | 4.3828e-3 | 2.0548e-3 | 8.4138e-4 | 7.9466e-5 |
| Loss : 42.9 dB | $e_{bit,D_0D_1}^{X}$ | $Q_{\alpha\alpha,D_0D_1}^{X}$ | $Y_{00,D_0D_1}^{U}$ | $Y_{11,D_0D_1}^{U}$ | $Y_{02,D_0D_1}^{U}$ | $Y_{20,D_0D_1}^{U}$ | $e_{ph,D_0D_1}^{U}$ | $R_{D_0D_1}$ |
| $D_0=1$, $D_1=0$ | 2.9411e-2 | 2.4285e-4 | 3.3624e-7 | 1.4031e-3 | 1.9558e-3 | 3.5447e-3 | 1.5101e-2 | 4.0372e-5 |
| $D_0=0$, $D_1=1$ | 2.4193e-2 | 2.6956e-4 | 2.4218e-7 | 1.3445e-3 | 1.5917e-3 | 2.2155e-3 | 1.0421e-2 | 4.8908e-5 |
| Loss : 42.9 dB | $e_{bit,D_0D_1}^{X}$ | $Q_{\alpha\alpha,D_0D_1}^{X}$ | $Y_{00,D_0D_1}^{U}$ | $Y_{11,D_0D_1}^{U}$ | $Y_{02,D_0D_1}^{U}$ | $Y_{20,D_0D_1}^{U}$ | $e_{ph,D_0D_1}^{U}$ | $R_{D_0D_1}$ |
| $D_0=1$, $D_1=0$ | 2.9411e-2 | 2.4285e-4 | 4.2824e-7 | 1.4466e-3 | 1.8264e-3 | 2.3963e-3 | 1.4481e-2 | 4.0601e-5 |

| $D_0 = 0$, $D_1 = 1$ | 2.4193e-2 | 2.6956e-4 | 3.9030e-7 | 1.1049e-3 | 1.9502e-3 | 2.3483e-3 | 1.2455e-2 | 4.8026e-5 |
|---|---|---|---|---|---|---|---|---|
| Loss : 42.9 dB | $e^X_{bit,D_0D_1}$ | $Q^X_{\alpha\alpha,D_0D_1}$ | $Y^U_{00,D_0D_1}$ | $Y^U_{11,D_0D_1}$ | $Y^U_{02,D_0D_1}$ | $Y^U_{20,D_0D_1}$ | $e^U_{ph,D_0D_1}$ | $R_{D_0D_1}$ |
| $D_0 = 1$, $D_1 = 0$ | 2.9411e-2 | 2.4285e-4 | 1.8107e-7 | 1.2939e-3 | 1.9913e-3 | 1.8607e-3 | 1.0792e-2 | 4.2009e-5 |
| $D_0 = 0$, $D_1 = 1$ | 2.4193e-2 | 2.6956e-4 | 1.4872e-7 | 1.0281e-3 | 1.7302e-3 | 2.2563e-3 | 9.1768e-3 | 4.9466e-5 |
| Loss : 52.9 dB | $e^X_{bit,D_0D_1}$ | $Q^X_{\alpha\alpha,D_0D_1}$ | $Y^U_{00,D_0D_1}$ | $Y^U_{11,D_0D_1}$ | $Y^U_{02,D_0D_1}$ | $Y^U_{20,D_0D_1}$ | $e^U_{ph,D_0D_1}$ | $R_{D_0D_1}$ |
| $D_0 = 1$, $D_1 = 0$ | 3.2258e-2 | 5.2524e-5 | 9.5721e-8 | 2.0381e-4 | 6.7083e-4 | 4.2181e-4 | 2.0380e-3 | 9.3901e-6 |
| $D_0 = 0$, $D_1 = 1$ | 2.7522e-2 | 5.2403e-5 | 1.0032e-7 | 3.6383e-4 | 7.4231e-4 | 1.0947e-3 | 2.2798e-3 | 9.5865e-6 |
| Loss : 52.9 dB | $e^X_{bit,D_0D_1}$ | $Q^X_{\alpha\alpha,D_0D_1}$ | $Y^U_{00,D_0D_1}$ | $Y^U_{11,D_0D_1}$ | $Y^U_{02,D_0D_1}$ | $Y^U_{20,D_0D_1}$ | $e^U_{ph,D_0D_1}$ | $R_{D_0D_1}$ |
| $D_0 = 1$, $D_1 = 0$ | 3.2258e-2 | 5.2524e-5 | 6.7879e-8 | 3.5241e-4 | 6.1979e-4 | 1.0352e-3 | 4.8327e-3 | 9.4199e-6 |
| $D_0 = 0$, $D_1 = 1$ | 2.7522e-2 | 5.2403e-5 | 5.8258e-8 | 2.2291e-4 | 4.7625e-4 | 9.7137e-4 | 4.3824e-3 | 9.8084e-6 |
| Loss : 52.9 dB | $e^X_{bit,D_0D_1}$ | $Q^X_{\alpha\alpha,D_0D_1}$ | $Y^U_{00,D_0D_1}$ | $Y^U_{11,D_0D_1}$ | $Y^U_{02,D_0D_1}$ | $Y^U_{20,D_0D_1}$ | $e^U_{ph,D_0D_1}$ | $R_{D_0D_1}$ |
| $D_0 = 1$, $D_1 = 0$ | 3.2258e-2 | 5.2524e-5 | 9.2309e-8 | 2.3626e-4 | 5.0173e-4 | 6.3143e-4 | 4.9598e-3 | 9.4071e-6 |
| $D_0 = 0$, $D_1 = 1$ | 2.7522e-2 | 5.2403e-5 | 8.7143e-8 | 1.5081e-4 | 8.1823e-4 | 1.0004e-3 | 5.8666e-3 | 9.6575e-6 |

Values of parameters $Q^X_{\alpha\alpha,D_0D_1}$, $e^X_{bit,D_0D_1}$, $Y^U_{nm,D_0D_1}$, $e^U_{ph,D_0D_1}$, and $R_{D_0D_1}$, where ($D_0$, $D_1$) ∈ {(1,0), (0,1)} and ($n$, $m$) ∈ {(0,0), (0,2), (2,0), (1,1)}, for three different total system loss values: 33.7 dB, 42.9 dB, and 52.9 dB.

**Table S2. Experimentally observed signal gains**

| Loss : 33.7 dB | $Q^{\mu_a,\mu_b}_{D_0,D_1}$ | $Q^{\mu_a,\nu_b}_{D_0,D_1}$ | $Q^{\nu_a,\mu_b}_{D_0,D_1}$ | $Q^{\nu_a,\nu_b}_{D_0,D_1}$ | $Q^{\omega_a,\omega_b}_{D_0,D_1}$ | $Q^{\mu_a,\omega_b}_{D_0,D_1}$ | $Q^{\omega_a,\mu_b}_{D_0,D_1}$ | $Q^{\nu_a,\omega_b}_{D_0,D_1}$ | $Q^{\omega_a,\nu_b}_{D_0,D_1}$ |
|---|---|---|---|---|---|---|---|---|---|
| $D_0 = 1$, $D_1 = 0$ | 2.7834e-4 | 1.5936e-4 | 1.5263e-4 | 3.6421e-5 | 1.4210e-6 | 1.4039e-4 | 1.3957e-4 | 1.8719e-5 | 1.7894e-5 |
| D0 = 0, D1 = 1 | 2.8294e-4 | 1.6218e-4 | 1.6223e-4 | 3.5892e-5 | 1.5263e-6 | 1.3647e-4 | 1.4211e-4 | 1.8210e-5 | 1.7526e-5 |
| Loss : 33.7 dB | $Q^{\mu_a,\mu_b}_{D_0,D_1}$ | $Q^{\mu_a,\nu_b}_{D_0,D_1}$ | $Q^{\nu_a,\mu_b}_{D_0,D_1}$ | $Q^{\nu_a,\nu_b}_{D_0,D_1}$ | $Q^{\omega_a,\omega_b}_{D_0,D_1}$ | $Q^{\mu_a,\omega_b}_{D_0,D_1}$ | $Q^{\omega_a,\mu_b}_{D_0,D_1}$ | $Q^{\nu_a,\omega_b}_{D_0,D_1}$ | $Q^{\omega_a,\nu_b}_{D_0,D_1}$ |
| D0 = 1, D1 = 0 | 2.7802e-4 | 1.6658e-4 | 1.5721e-4 | 3.7258e-5 | 1.7927e-6 | 1.3764e-4 | 1.3630e-4 | 1.8521e-5 | 1.8267e-5 |
| D0 = 0, D1 = 1 | 2.8778e-4 | 1.5193e-4 | 1.6309e-4 | 3.5669e-5 | 1.6282e-6 | 1.3323e-4 | 1.4218e-4 | 1.8556e-5 | 1.8483e-5 |
| Loss : 33.7 dB | $Q^{\mu_a,\mu_b}_{D_0,D_1}$ | $Q^{\mu_a,\nu_b}_{D_0,D_1}$ | $Q^{\nu_a,\mu_b}_{D_0,D_1}$ | $Q^{\nu_a,\nu_b}_{D_0,D_1}$ | $Q^{\omega_a,\omega_b}_{D_0,D_1}$ | $Q^{\mu_a,\omega_b}_{D_0,D_1}$ | $Q^{\omega_a,\mu_b}_{D_0,D_1}$ | $Q^{\nu_a,\omega_b}_{D_0,D_1}$ | $Q^{\omega_a,\nu_b}_{D_0,D_1}$ |
| D0 = 1, D1 = 0 | 2.7770e-4 | 1.6154e-4 | 1.5430e-4 | 3.6821e-5 | 1.5789e-6 | 1.4045e-4 | 1.3760e-4 | 1.9231e-5 | 1.8046e-5 |
| D0 = 0, D1 = 1 | 2.7948e-4 | 1.5568e-4 | 1.6226e-4 | 3.5465e-5 | 1.4777e-6 | 1.3762e-4 | 1.4164e-4 | 1.8623e-5 | 1.8238e-5 |
| Loss : 42.9 dB | $Q^{\mu_a,\mu_b}_{D_0,D_1}$ | $Q^{\mu_a,\nu_b}_{D_0,D_1}$ | $Q^{\nu_a,\mu_b}_{D_0,D_1}$ | $Q^{\nu_a,\nu_b}_{D_0,D_1}$ | $Q^{\omega_a,\omega_b}_{D_0,D_1}$ | $Q^{\mu_a,\omega_b}_{D_0,D_1}$ | $Q^{\omega_a,\mu_b}_{D_0,D_1}$ | $Q^{\nu_a,\omega_b}_{D_0,D_1}$ | $Q^{\omega_a,\nu_b}_{D_0,D_1}$ |
| D0 = 1, D1 = 0 | 2.1645e-4 | 1.3597e-4 | 1.3747e-4 | 5.3373e-5 | 1.6131e-6 | 1.1252e-4 | 1.1063e-4 | 2.2176e-5 | 2.6447e-5 |
| D0 = 0, D1 = 1 | 2.1077e-4 | 1.3644e-4 | 1.3050e-4 | 5.2365e-5 | 1.7184e-6 | 1.1324e-4 | 1.0701e-4 | 2.6463e-5 | 2.6634e-5 |
| Loss : 42.9 dB | $Q^{\mu_a,\mu_b}_{D_0,D_1}$ | $Q^{\mu_a,\nu_b}_{D_0,D_1}$ | $Q^{\nu_a,\mu_b}_{D_0,D_1}$ | $Q^{\nu_a,\nu_b}_{D_0,D_1}$ | $Q^{\omega_a,\omega_b}_{D_0,D_1}$ | $Q^{\mu_a,\omega_b}_{D_0,D_1}$ | $Q^{\omega_a,\mu_b}_{D_0,D_1}$ | $Q^{\nu_a,\omega_b}_{D_0,D_1}$ | $Q^{\omega_a,\nu_b}_{D_0,D_1}$ |
| D0 = 1, D1 = 0 | 2.1378e-4 | 1.3546e-4 | 1.3458e-4 | 5.3813e-5 | 1.8605e-6 | 1.1189e-4 | 1.1013e-4 | 2.5650e-5 | 2.6886e-5 |
| D0 = 0, D1 = 1 | 2.0770e-4 | 1.3738e-4 | 1.2921e-4 | 5.1644e-5 | 1.8131e-6 | 1.1443e-4 | 1.0719e-4 | 2.6455e-5 | 2.5673e-5 |
| Loss : 42.9 dB | $Q^{\mu_a,\mu_b}_{D_0,D_1}$ | $Q^{\mu_a,\nu_b}_{D_0,D_1}$ | $Q^{\nu_a,\mu_b}_{D_0,D_1}$ | $Q^{\nu_a,\nu_b}_{D_0,D_1}$ | $Q^{\omega_a,\omega_b}_{D_0,D_1}$ | $Q^{\mu_a,\omega_b}_{D_0,D_1}$ | $Q^{\omega_a,\mu_b}_{D_0,D_1}$ | $Q^{\nu_a,\omega_b}_{D_0,D_1}$ | $Q^{\omega_a,\nu_b}_{D_0,D_1}$ |
| D0 = 1, D1 = 0 | 2.1509e-4 | 1.3473e-4 | 1.3781e-4 | 5.2152e-5 | 1.6833e-6 | 1.1091e-4 | 1.1258e-4 | 2.6884e-5 | 2.6936e-5 |
| D0 = 0, D1 = 1 | 2.0981e-4 | 1.3784e-4 | 1.3259e-4 | 5.1856e-5 | 1.6208e-6 | 1.1517e-4 | 1.0679e-4 | 2.6828e-5 | 2.6084e-5 |
| Loss : 52.9 dB | $Q^{\mu_a,\mu_b}_{D_0,D_1}$ | $Q^{\mu_a,\nu_b}_{D_0,D_1}$ | $Q^{\nu_a,\mu_b}_{D_0,D_1}$ | $Q^{\nu_a,\nu_b}_{D_0,D_1}$ | $Q^{\omega_a,\omega_b}_{D_0,D_1}$ | $Q^{\mu_a,\omega_b}_{D_0,D_1}$ | $Q^{\omega_a,\mu_b}_{D_0,D_1}$ | $Q^{\nu_a,\omega_b}_{D_0,D_1}$ | $Q^{\omega_a,\nu_b}_{D_0,D_1}$ |
| D0 = 1, D1 = 0 | 3.9324e-5 | 2.2302e-5 | 2.2994e-5 | 4.0236e-6 | 2.2105e-7 | 2.1118e-5 | 2.2668e-5 | 2.1052e-6 | 2.1394e-6 |
| D0 = 0, D1 = 1 | 4.0203e-5 | 2.2060e-5 | 2.2436e-5 | 3.6447e-6 | 2.1578e-7 | 2.2010e-5 | 2.1789e-5 | 1.8578e-6 | 2.0078e-6 |

| Loss : 52.9 dB | $Q^{\mu_a,\mu_b}_{D_0,D_1}$ | $Q^{\mu_a,\nu_b}_{D_0,D_1}$ | $Q^{\nu_a,\mu_b}_{D_0,D_1}$ | $Q^{\nu_a,\nu_b}_{D_0,D_1}$ | $Q^{\omega_a,\omega_b}_{D_0,D_1}$ | $Q^{\mu_a,\omega_b}_{D_0,D_1}$ | $Q^{\omega_a,\mu_b}_{D_0,D_1}$ | $Q^{\nu_a,\omega_b}_{D_0,D_1}$ | $Q^{\omega_a,\nu_b}_{D_0,D_1}$ |
|---|---|---|---|---|---|---|---|---|---|
| D0 = 1, D1 = 0 | 3.8268e-5 | 2.2810e-5 | 2.2717e-5 | 3.9052e-6 | 1.7894e-7 | 2.1561e-5 | 2.0997e-5 | 1.8131e-6 | 1.9605e-6 |
| D0 = 0, D1 = 1 | 3.9879e-5 | 2.2905e-5 | 2.2489e-5 | 3.6421e-6 | 1.6578e-7 | 2.0061e-5 | 2.0152e-5 | 1.6842e-6 | 1.9394e-6 |
| Loss : 52.9 dB | $Q^{\mu_a,\mu_b}_{D_0,D_1}$ | $Q^{\mu_a,\nu_b}_{D_0,D_1}$ | $Q^{\nu_a,\mu_b}_{D_0,D_1}$ | $Q^{\nu_a,\nu_b}_{D_0,D_1}$ | $Q^{\omega_a,\omega_b}_{D_0,D_1}$ | $Q^{\mu_a,\omega_b}_{D_0,D_1}$ | $Q^{\omega_a,\mu_b}_{D_0,D_1}$ | $Q^{\nu_a,\omega_b}_{D_0,D_1}$ | $Q^{\omega_a,\nu_b}_{D_0,D_1}$ |
| D0 = 1, D1 = 0 | 3.7688e-5 | 2.2367e-5 | 2.2700e-5 | 3.9909e-6 | 2.1714e-7 | 2.2170e-5 | 2.1673e-5 | 2.1058e-6 | 2.1176e-6 |
| D0 = 0, D1 = 1 | 2.8461e-5 | 2.2229e-5 | 2.2421e-5 | 3.8409e-6 | 1.9714e-7 | 2.1050e-5 | 2.2155e-5 | 1.7911e-6 | 1.9941e-6 |

Experimentally observed signal gains $Q^{D_0,D_1}_{\mu_a,\mu_b}$ in the decoy mode for three different total channel loss values (33.7 dB, 42.9 dB, and 52.9 dB), under the conditions $(\mu_a, \mu_b) \in \{\mu, \nu, \omega\}$ and $(D_0, D_1) \in \{(1,0), (0,1)\}$.

**Supplementary Note S7: Multi-Sideband Modulation and Encoding**

This section elaborates on an important extension of the SSB-TF-QKD protocol, namely from single-sideband to multi-sideband interference, and presents our experimental verification of the key multi-sideband modulation technology. This extension retains the core advantage of the scheme, eliminating the need for global phase locking, while potentially doubling the encoding efficiency and thus further enhancing the secure key rate.

As described in Supplementary Note S1, when the bias voltage of the I/Q modulator is properly set, the output optical field is a superposition of the carrier and a single sideband. From a more general perspective, by adjusting three key parameters of the modulator, the phase difference $\Delta\varphi$ between the I and Q RF signals and the DC bias phase $\varphi_{DC3}$ of the modulator, both the positive and negative first-order sidebands can be generated simultaneously, with independently controllable amplitudes and phases.

From Eq. (S6) in Supplementary Note S1, the output optical field after the I/Q modulator can be expanded as a superposition of sidebands of various orders. For the positive and negative first-order sidebands, their complex amplitudes can be written respectively as:

$$E_{+1} \propto 2i \cdot \cos\left(\left(\Delta\varphi + \varphi_{DC3}\right)/2\right) \cdot e^{i\left[(\omega_c+\Omega)t+\theta t+\varphi_A^{RF}+(\Delta\varphi+\varphi_{DC3})/2\right]}, \tag{S34}$$

$$E_{-1} \propto 2i \cdot \cos\left(\left(\Delta\varphi - \varphi_{DC3}\right)/2\right) \cdot e^{i\left[(\omega_c+\Omega)t+\theta t-\varphi_A^{RF}-(\Delta\varphi-\varphi_{DC3})/2\right]}, \tag{S35}$$

where $\Delta\varphi$ is the phase difference between the I and Q RF signals, $\varphi_{DC3}$ is the bias phase of the modulator, and $\varphi_A^{RF}$ is the reference phase of the RF driving signal. The above expressions reveal an important fact, by independently adjusting $\Delta\varphi$ and $\varphi_{DC3}$, the amplitudes and phases of the positive and negative first-order sidebands can be controlled independently. Specifically, the amplitude of the positive first-order sideband is determined by $\cos[(\Delta\varphi+\varphi_{DC3})/2]$, and its phase is given by $\varphi_A^{RF} + \left(\Delta\varphi + \varphi_{DC3}\right)/2 + \pi/2$; the amplitude of the negative first-order sideband is determined by $\cos[(\Delta\varphi-\varphi_{DC3})/2]$, and its phase is given by $-\varphi_A^{RF} - \left(\Delta\varphi - \varphi_{DC3}\right)/2 + \pi/2$.

This physical mechanism provides a theoretical foundation for the extension of the proposed scheme. For a single optical pulse, the positive and negative first-order sidebands can be simultaneously utilized as two independent quantum signal, each carrying different key information, as shown in Table S3.

**Table S3. Phase parameters and sideband amplitudes for implementing multi-sideband encoding.**

| Code | $\varphi_A^{RF}$ | $\Delta\varphi$ | $\varphi_{DC3}$ | $\cos\left(\frac{\Delta\varphi+\varphi_{DC3}}{2}\right)$ | $\cos\left(\frac{\Delta\varphi-\varphi_{DC3}}{2}\right)$ |
|---|---|---|---|---|---|
| (0, 0) | 0 | 0 | 0 | 1 | 1 |
| (π, π) | 0 | 0 | 0 | 1 | 1 |

| (0, π) | 0 | -π | π | 1 | 1 |
|---|---|---|---|---|---|
| (π, 0) | 0 | π | π | 1 | 1 |

Here, $\varphi_A^{RF}$ is the reference phase of the RF signal at the transmitting end, $\Delta\varphi$ is the phase difference between the I and Q signals applied to the modulator, $\varphi_{DC3}$ is the bias phase applied to the modulator, $\cos[(\Delta\varphi+\varphi_{DC3})/2]$ is the amplitude coefficient of the positive first-order sideband, and $\cos[(\Delta\varphi-\varphi_{DC3})/2]$ is the amplitude coefficient of the negative first-order sideband. As shown in the table, under the four encoding combinations, the amplitude coefficients of the positive and negative first-order sidebands are equal, satisfying the security requirements in QKD.

To verify the feasibility of multi-sideband modulation, we built a test system and experimentally characterized the key parameters. Figure S8 shows a typical multi-sideband modulation spectrum. By setting parameter combinations such as $\Delta\varphi=\pi/2$ and $\varphi_{DC}=\pi/2$, equal-amplitudes of both the positive and negative first-order sidebands were successfully achieved. In the spectrum, the carrier, the positive and negative first-order sidebands are clearly visible, with their intensities being essentially equal. The suppression ratios of both the sidebands relative to unwanted higher-order sidebands (e.g., ±2 order) exceed 30 dB, indicating good modulation quality and the negligible impact of higher-order components on the quantum signal.

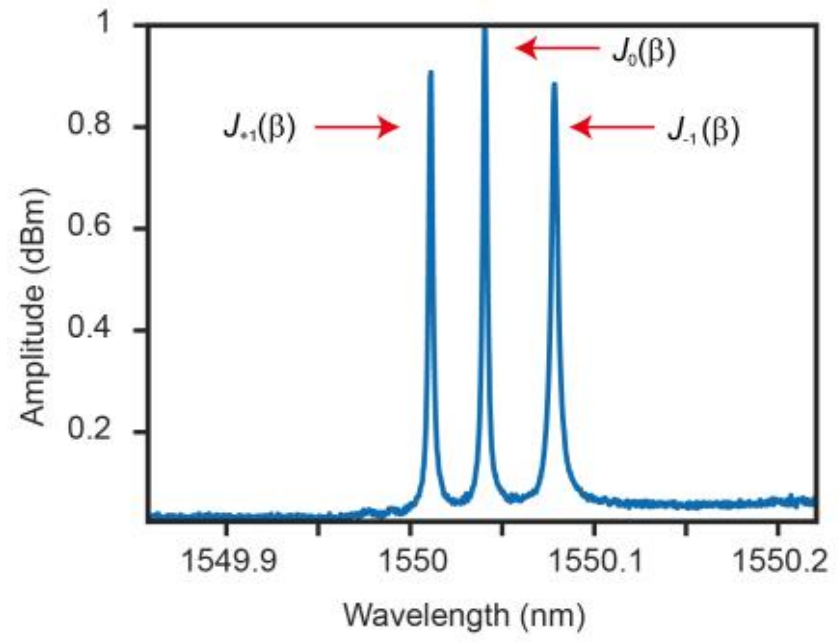

**Figure S8. Measured spectrum of multi-sideband modulation.**

Extending this scheme to multi-sideband interference offers several advantages. First, the encoding efficiency increases from 1 bit per pulse to 2 bits per pulse. Under the same system resources and channel conditions, the secure key rate can theoretically be nearly doubled. Second, it makes full use of the modulation bandwidth of the I/Q modulator, increasing the key generation rate without raising the pulse repetition frequency. Third, it allows dynamic switching between single-sideband and multi-sideband modes based on channel conditions and user requirements, enabling a trade-off between key rate and robustness.